\documentclass[english,aps,showpacs,superscriptaddress,tightenlines,twocolumn,pra]{revtex4}

\usepackage[T1]{fontenc}
\usepackage[utf8]{inputenc}
\usepackage{babel}
\usepackage{amsmath}
\usepackage{amssymb}
\usepackage{braket}
\usepackage{nicefrac}
\usepackage{tikz}
\usetikzlibrary{arrows,decorations.pathmorphing,decorations.markings,decorations.shapes,backgrounds,positioning,fit,trees,calc} 

\tikzset{snake it/.style={decorate, decoration={snake,amplitude=.4mm,segment length=2mm,pre length=1mm,post length=1mm}}}
\tikzset{
   vertex/.style={circle, inner sep=0pt, minimum size=5pt,fill=black,label=#1}, vertex/.default=\text{},
   crossing/.style={circle, inner sep=0, minimum size=0,label=#1}, crossing/.default=\text{},
   named/.style={draw,circle, inner sep=0pt, minimum size=12pt},
   namedE/.style={draw,circle, inner sep=0pt, minimum size=7pt},
   bath/.style={draw,thick,->-=.5}, bath/.default={},
   bathr/.style={draw,thick,dashed,->-=.5},
   time/.style={draw,dashed,thin},
   system/.style={draw,thick}, system/.default={},
   syslabel/.style={midway,auto,green!40!black}}
   \tikzset{dotted pattern/.style args={#1 and #2}{
   decorate,
   fill=black,
   decoration={
    shape backgrounds,
    shape=circle,
    shape size=#1,
    shape sep={#2, between center},
    }
  },
  dotted pattern/.default={1pt and 1.5mm},
}

\definecolor{kit-green100}{cmyk}{1.00,0.00,0.60,0.00}
\colorlet{kit-green70}{kit-green100!70}
\colorlet{kit-green50}{kit-green100!50}
\colorlet{kit-green30}{kit-green100!30}
\colorlet{kit-green15}{kit-green100!15}

\definecolor{kit-blue100}{cmyk}{0.80,0.50,0.00,0.00}
\colorlet{kit-blue70}{kit-blue100!70}
\colorlet{kit-blue50}{kit-blue100!50}
\colorlet{kit-blue30}{kit-blue100!30}
\colorlet{kit-blue15}{kit-blue100!15}

\definecolor{kit-palegreen100}{cmyk}{.6,0,1,0}
\colorlet{kit-palegreen70}{kit-palegreen100!70}
\colorlet{kit-palegreen50}{kit-palegreen100!50}
\colorlet{kit-palegreen30}{kit-palegreen100!30}
\colorlet{kit-palegreen15}{kit-palegreen100!15}

\definecolor{kit-yellow100}{cmyk}{0.00,0.05,1.00,0.00}
\colorlet{kit-yellow70}{kit-yellow100!70}
\colorlet{kit-yellow50}{kit-yellow100!50}
\colorlet{kit-yellow30}{kit-yellow100!30}
\colorlet{kit-yellow15}{kit-yellow100!15}

\definecolor{kit-orange100}{cmyk}{0.00,0.45,1.00,0.00}
\colorlet{kit-orange70}{kit-orange100!70}
\colorlet{kit-orange50}{kit-orange100!50}
\colorlet{kit-orange30}{kit-orange100!30}
\colorlet{kit-orange15}{kit-orange100!15}

\definecolor{kit-brown100}{cmyk}{.35,.5,1,0}
\colorlet{kit-brown70}{kit-brown100!70}
\colorlet{kit-brown50}{kit-brown100!50}
\colorlet{kit-brown30}{kit-brown100!30}
\colorlet{kit-brown15}{kit-brown100!15}

\definecolor{kit-red100}{cmyk}{.25,1,1,0}
\colorlet{kit-red70}{kit-red100!70}
\colorlet{kit-red50}{kit-red100!50}
\colorlet{kit-red30}{kit-red100!30}
\colorlet{kit-red15}{kit-red100!15}

\definecolor{kit-lilac100}{cmyk}{.25,1,0,0}
\colorlet{kit-lilac70}{kit-lilac100!70}
\colorlet{kit-lilac50}{kit-lilac100!50}
\colorlet{kit-lilac30}{kit-lilac100!30}
\colorlet{kit-lilac15}{kit-lilac100!15}

\definecolor{kit-cyanblue100}{cmyk}{.9,.05,0,0}
\colorlet{kit-cyanblue70}{kit-cyanblue100!70}
\colorlet{kit-cyanblue50}{kit-cyanblue100!50}
\colorlet{kit-cyanblue30}{kit-cyanblue100!30}
\colorlet{kit-cyanblue15}{kit-cyanblue100!15}
\begin{document}

\title{Reconstructing the ideal results of a perturbed analog quantum simulator}

\author{Iris Schwenk}
\affiliation{Institute of Theoretical Solid State Physics,
      Karlsruhe Institute of Technology (KIT), 76131 Karlsruhe, Germany}

\author{Jan-Michael Reiner}
\affiliation{Institute of Theoretical Solid State Physics,
      Karlsruhe Institute of Technology (KIT), 76131 Karlsruhe, Germany}

\author{Sebastian Zanker}
\affiliation{Institute of Theoretical Solid State Physics,
      Karlsruhe Institute of Technology (KIT), 76131 Karlsruhe, Germany}

\author{Lin Tian}
\affiliation{School of Natural Sciences, University of California, Merced, California 95343, USA}    

\author{Juha Lepp\"akangas}
\affiliation{Institute of Theoretical Solid State Physics,
      Karlsruhe Institute of Technology (KIT), 76131 Karlsruhe, Germany}

\author{Michael Marthaler}
\affiliation{Institut für Theorie der Kondensierten Materie (TKM), Karlsruhe Institute of Technology (KIT), 76131 Karlsruhe, Germany}
\affiliation{Theoretical Physics, Saarland University, 66123 Saarbrücken, Germany}

\date{\today}

\begin{abstract}
Well-controlled quantum systems can potentially be used as quantum simulators. However, a quantum simulator is inevitably perturbed by coupling to additional degrees of freedom.
This constitutes a major roadblock to useful quantum simulations. So far there are only limited means to understand the effect of perturbation on the results of quantum simulation.
Here, we present a method which, in certain circumstances, allows for the reconstruction of the ideal result from measurements on a perturbed quantum simulator.
We consider extracting the value of the correlator $\braket{\hat{O}^i(t) \hat{O}^j(0)}$ from the simulated system, where $\hat{O}^i$ are the operators which couple the system to its environment. The ideal correlator can be straightforwardly reconstructed by using statistical knowledge of the environment, if any $n$-time correlator of operators $\hat{O}^i$ of the ideal system can be written as products of two-time correlators.
We give an approach to verify the validity of this assumption experimentally by additional measurements on the perturbed quantum simulator.
The proposed method can allow for reliable quantum simulations with systems subjected to environmental noise without adding an overhead to the quantum system.
\end{abstract}


\maketitle

\section{Introduction and central results}\label{sec_Introduction_and_central_results}

Today we possess in principle the full knowledge to describe all processes of interest in 
a wide range of fields, such as chemistry, biology and solid state physics. 
In all these fields a truly microscopic description is  possible using quantum mechanics.
However, it is also well understood that in practice full quantum mechanical simulations of even modestly-sized systems are impossible~\cite{Troyer_1}. 
To efficiently study quantum problems, we need to use other, 
well controlled quantum mechanical systems~\cite{Cold_Gases_Simulator,Ion_Simulator,Cirac_Review,Quantum_Simulator_EPJ}.
In recent years unprecedented direct control over quantum systems has been 
achieved~\cite{Super_Review,Ion_Review,Rydberg_Review,Marinis_Threshold,Ion_Simulator}.
Precise experiments in the quantum regime have been performed using atomic systems~\cite{Trapped_Fermi_2002,Trapped_Fermi_2006,Frustrated_Spins}, 
superconducting qubits~\cite{IBM_Fault_tolerance,DiCarlo_Stabilized,Devoret_Cat_states,Wallraff_Spins,Martinis_Quantum_Simulations}, 
photonic circuits~\cite{Photonic_Hydrogen,Boson_Sampling_1,Boson_Sampling_2},
and nuclear spins~\cite{Nuclear_Wrachtrup,Nuclear_Phosphor}.
Larger systems have been demonstrated using trapped ions~\cite{350_Spin_simulator,Bohnet2016} and the equilibration of interacting bosons has been studied in cold gases~\cite{Bloch_Equilbiration,Eisert_Equlibiration}.

A promising approach to understanding quantum systems is analog quantum simulation~\cite{Analog_Manousakis}, where the goal is to create an artificial system with a Hamiltonian that is equivalent to the system we intend to study. 
Apart from quantum simulations using cold gases~\cite{Cirac_Cold_gases,Fermi_Sea_Heidelberg} and trapped ions~\cite{Cirac_PRL,Quantum_Magnet_Schaetz}, there are many proposals for analog quantum simulation with superconducting circuits~\cite{Photosynthesis_Super,Tian_Simu_1,Tian_Simu_2,Solan_digital_analog}, exploiting the controllability of superconducting systems, which in principle allows the creation of a large class of Hamiltonians. 
While most current superconducting systems are relatively small~\cite{Fermi_Simulation_super_qubits,Weak_localized_super_qubits}, 
larger networks of superconducting non-linear elements are now being explored~\cite{Pascal_Meta,D_Wave500Qubits,D_Wave1000_qubits}.
Other architectures for analog quantum simulation have also been investigated~\cite{Nature_Plenio,Nature_Silicon}.

In this article, we study an analog quantum simulator with the ideal Hamiltonian $H_S$. 
To understand the properties of the simulated system, we would like to use a measurement to extract a time-ordered correlation function (Green's function),
\begin{eqnarray}\label{eq_Ideal_Correlator} 
iG_{S0}(t) &=& \langle {\cal T} \hat{O}(t)\hat{O}(0) \rangle_0\\
             &=& \bra{0} {\cal T} e^{i H_S t}\hat{O} e^{-i H_S t} \hat{O} \ket{0}\, , \nonumber
\end{eqnarray}
where ${\cal T}$ is the time-ordering operator. 
The index $S0$ indicates that we are considering the ideal Green's function of the unperturbed Hamiltonian $H_S$, without coupling to additional degrees of freedom, and $\ket{0}$ is the ground state of $H_S$ (zero-temperature limit).
We start our analysis from this simple example and later in Sec.~\ref{sec_Full_model_and_discussion}
extend the theory to multiple operators $\hat{O}^i$, and to finite temperatures. 
We consider time-ordered Green's functions, since these are in general connected  to numerous quantities of interest in experiments,  such as  heat or electric transport coefficients.
There are several proposals which describe methods to measure the relevant correlators in the context of analog quantum simulation~\cite{Tian_Correlator,Tian_ReadoutReconstruction,Correlator_Buchleitner,Correlator_Anderes_PAper}. 
Thus, we assume that Green's functions play a central role in extracting results from a quantum simulator.

However, if we want to use measurements on a quantum simulator to study the properties of an ideal Hamiltonian,
the key challenge remains: What is the role of errors and imperfections of the artificial system in a real measurement~\cite{Quantum_Simulator_EPJ,Can_we_trust_Emulator,Reliability_AQS,Certification_Eisert}?
Usually we quantify the influence of external degrees of freedom by comparing measurements to theoretical predictions. 
However, by definition, for quantum simulation it should not be possible to predict the result; neither analytically nor numerically using classical computers.
Some proposals exist to analyze \cite{IrisPRL} or mitigate \cite{mitigationLi,mitigationGambetta} errors for small noise in analog or digital quantum simulators.
The approach we introduce in this paper works potentially also for intermediate levels of noise strength. It is based on connecting the ideal Green's function, Eq.~(\ref{eq_Ideal_Correlator}), to the perturbed Green's function we measure using a quantum simulator.
We consider Green's functions where $\hat{O}$ is also the operator by which the quantum simulator couples to additional degrees of freedom (which cause the errors).
This restricts the generality of the approach, but in reality it is actually very likely that the same mechanism which connects the system to its bath also allows for the readout of the system.
For example, readout via a resonator for modern superconducting qubits can be done dispersively (via $\sigma_z$) or resonantly (via $\sigma_x$). In the case of $T_1$ limited qubits with resonant readout or $T_2$ limited qubits with dispersive readout \cite{IBMqubits} our requirement is fulfilled.
So it is reasonable to assume that this is one of the Green's functions to which we have an easy access in experiments. 

We show that under specific conditions it is in fact possible to extract the ideal correlator of the operator $\hat{O}$ even from a perturbed system. 
One ingredient in our approach is a good statistical knowledge of the 
additional degrees of freedom which act on $H_S$. 
This assumption is justified, for example, for a quantum simulator build from tunable qubits,
where qubits can be decoupled and the properties of the baths of individual qubits can be probed by established spectroscopical methods.
Apart from this, only one assumption is necessary about the properties of the ideal correlators.
We need that any $n$-time correlation function can be expressed as the product of
two-time correlation functions. 
This condition will be discussed in more detail in Sec.~\ref{sec:PrincipalIdea}.
In the present paper we describe this method assuming $\hat{O}$ and that the additional degrees of freedom are bosonic, but the method can also be directly transferred to fermionic operators $\hat{O}$ and fermionic baths.

\subsection{Principal idea}\label{sec:PrincipalIdea}

We start by presenting a simple example of our approach, where we show how to extract the ideal 
properties from an imperfect simulator in equilibrium.
In Sec.~\ref{sec_Full_model_and_discussion}, we extend this result to more general situations.

The full system  we consider can be described by the  Hamiltonian,
\begin{equation}\label{eq_Full_Hamiltonian}
 H=H_S+H_C+H_B\,\, ,\,\, H_C=\hat{O}\hat{X} \;.
\end{equation}
Here the ideal Hamiltonian of the simulator $H_S$ is coupled via the Hamiltonian $H_C$ to the
additional degrees of freedom contained in the bath Hamiltonian $H_B$. 
The system operator in $H_C$ is $\hat{O}$, which is the same as what we used to define the ideal correlator in Eq.~(\ref{eq_Ideal_Correlator}), and the bath operator is $\hat X$.

\begin{figure}[t]
\includegraphics{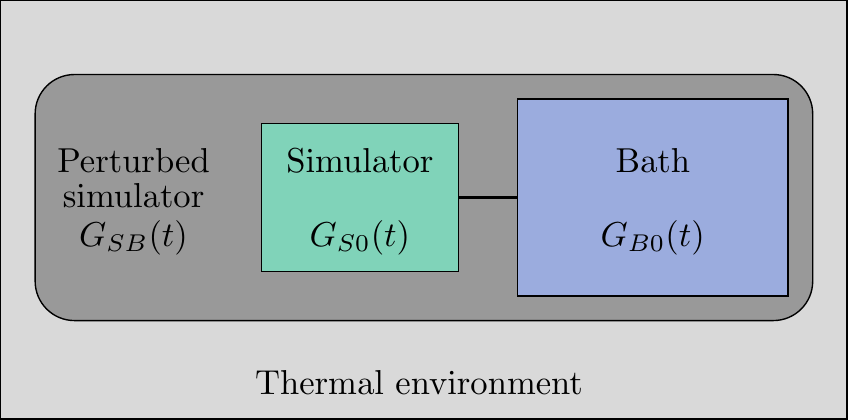}
\caption{The quantum simulator is coupled to a perturbative bath. The simulator-bath system is coupled weakly to an environment that establishes thermal equilibrium. For each sub-component of the system we define a free correlator: 
the ideal correlator of the simulator $iG_{S0}(t)=\langle {\cal T} \hat{O}(t)\hat{O}(0)\rangle_0$ as defined in Eq.~(\ref{eq_Ideal_Correlator}) and the free correlator
of the bath $iG_{B0}(t)=\langle{\cal T} \hat{X}(t)\hat{X}(0)\rangle_0$. 
The full correlator $iG_{SB}(t)=\langle{\cal T}  \hat{O}(t)\hat{O}(0)\rangle$ accounts for the coupling in the full Hamiltonian, Eq.~(\ref{eq_Full_Hamiltonian}).}\label{fig_total_system}
\end{figure}

The bath can usually be described by a set of bosonic modes and we assume that the free correlator of the bath $G_{B0}(t)$ is known, for example, from spectroscopic measurements. 
For the definition of all relevant Green's functions see Fig.~\ref{eq_Ideal_Correlator} and its caption.
In Sec.~\ref{sec_Full_model_and_discussion}, we give a more precise definition.

The total system described by $H$ is in thermal equilibrium. 
It should be emphasized that if coupling to the thermal bath is not infinitely weak it cannot be assumed that the only result of this coupling is the creation of equilibrium~ \cite{Strong_Coupling_Quantum,Strong_Coupling_Iris,Strong_Coupling_Classical}.
In the main part of this paper we focus on the situation at zero temperature and in Sec.~\ref{sec_finite_temperatures} extend our method to finite temperatures.

We want to connect the spectral function of the bath to the properties of the perturbed quantum simulator.
Standard many-body physics techniques exist which can be used to expand the full Green's function $G_{SB}(\omega)$ in terms of the ideal Green's functions $G_{S0}(\omega)$ and $G_{B0}(\omega)$~\cite{Greensfunction_Rickayzen}.
However, to apply these techniques there is one key assumption that is absolutely crucial:
Wick's theorem needs to apply in some form. Using this theorem it is possible to connect a single correlator of $2n$ operators with $n$ two-time correlators. 
Wick's theorem for the system operator $\hat{O}$ takes the form 
\begin{eqnarray}\label{eq_Wick_Theorem_For_O} 
  & &\langle {\cal T} \hat{O}(t_1)\hat{O}(t_2)\ldots\hat{O}(t_{n-1})\hat{O}(t_n) \rangle_0\\
  & &=
   \langle {\cal T} \hat{O}(t_1)\hat{O}(t_2) \rangle_0
   \langle {\cal T} \hat{O}(t_3)\ldots\hat{O}(t_{n-1})\hat{O}(t_n) \rangle_0 \nonumber \\
  & & + \langle {\cal T} \hat{O}(t_1)\hat{O}(t_3) \rangle_0
   \langle {\cal T} \hat{O}(t_2)\ldots\hat{O}(t_{n-1})\hat{O}(t_n) \rangle_0 \nonumber\\
     & & + \ldots + \langle {\cal T} \hat{O}(t_1)\hat{O}(t_n) \rangle_0
   \langle {\cal T} \hat{O}(t_2)\ldots\hat{O}(t_{n-1}) \rangle_0 \,. \nonumber
\end{eqnarray}
This relation can be applied repeatedly until only two-time correlators remain.
For the bath operator $\hat{X}$ it is natural to assume that Wick's theorem applies,
in accordance with numerous system-bath descriptions.
However, for the system operator $\hat{O}$ this is not in general true.
A well known case, where Eq.~(\ref{eq_Wick_Theorem_For_O}) holds is if the system $H_S$
can be described as a system of non-interacting quasiparticles and $\hat{O}$ can be written as a linear combination of the annihilation and creation operators of these quasiparticles. 
More generally, Eq.~(\ref{eq_Wick_Theorem_For_O}) is valid if the fluctuations of $\hat{O}(t)$ have a Gaussian distribution. 
The expansion of $n$-time correlators in pair and higher correlators has been studied extensively for spin systems~\cite{Wick_1,Wick_2,Wick_3} and deviations from 
Gaussian statistics have been studied in the field of full-counting statistics~\cite{Counting_1,Counting_2,Counting_3,Counting_4}.
From relatively general considerations, such as the central limit theorem~\cite{Central_Limit_Book}, we expect that fluctuations become more Gaussian as the system size increases, which is also the most interesting limit for a quantum simulator. 
However, in some systems non-Gaussian fluctuations are known to persist even at large system size~\cite{Non_Gaussian_Phase_transition,Non_Gaussian_Flow}
or become even size independent~\cite{Non_Gaussian_Resistance_Fluctuations_in_Disordered Materials}.
For different expansions it could also be useful to map qubits coupled to bosonic baths to an effective electron-phonon model\cite{basti}.
In Sec.~\ref{sec_Corrections_to_the_ Wick_Theorem} we discuss, how Eq.~(\ref{eq_Wick_Theorem_For_O}) can be checked, to some extend, by making appropriate measurements on the perturbed quantum simulator.

Assuming Eq.~(\ref{eq_Wick_Theorem_For_O}) holds, we find an exact relation between the Green's functions,
\begin{equation}\label{eq_Full_Equation_of_Correlator_Simplest_case}
 G_{SB}(\omega)=G_{S0}(\omega)+G_{S0}(\omega)G_{B0}(\omega)G_{SB}(\omega) \;.
\end{equation}
This is the well-known Dyson equation that defines the total Green's function
as a function of the free system and bath Green's functions.

\subsection{Central result}
From Eq.~(\ref{eq_Full_Equation_of_Correlator_Simplest_case}) we see that the perturbed quantum simulator can be used to find the correlator of the unperturbed simulator
$G_{S0}(\omega)$ as long as we know the free Green's function of the bath $G_{B0}(\omega)$, since
\begin{equation}\label{eq_Essential_Result_most_simple} 
 G_{S0}(\omega)\! = \!
 \frac{G_{SB}(\omega)}{1 + G_{B0}(\omega)G_{SB}(\omega)}\,.
\end{equation}
This states the central idea of this paper in the simplest form. 
To derive Eq.~(\ref{eq_Essential_Result_most_simple}) we use an important assumption:
that Wick's theorem in the form in Eq.~(\ref{eq_Wick_Theorem_For_O}) applies for the system operator $\hat{O}$.
This condition will be discussed in more detail in Sec.~\ref{sec_Corrections_to_the_ Wick_Theorem}, where we also show how to extract the lowest order correction to this result from the perturbed simulator.
Apart from this the quality of the reconstruction is also restricted by the precision of the knowledge of the correlators, which is the subject of Sec.~\ref{sec_imperfect_measurement}.
In particular, we presume that the properties of the bath are measured independently of the system, which will be discussed more detailed in Sec.~\ref{sec_Full_model_and_discussion}.
In Sec.~\ref{sec_Full_model_and_discussion}, we also consider the case where multiple baths couple to system via operators $\hat{O}^i$ and extend the reconstruction method to finite temperatures. 
Finally, we discuss a simple example, which can be solved analytically, to validate our result.

\section{Verifying Wick's Theorem}\label{sec_Corrections_to_the_ Wick_Theorem}

The validity of Wick's theorem for the system operator $\hat{O}$ is crucial for the  derivation of Eq.~(\ref{eq_Essential_Result_most_simple}); however, for non trivial systems we cannot in general predict if Wick's theorem holds. 
Therefore, we describe a method to verify the validity of Wick's theorem using the quantum simulator itself.
A detailed derivation is given in Appendix~\ref{app_4-time_correlator}. 

We introduce the lowest-order correction to Wick's theorem $G_4(t_1,t_2,t_3,t_4)$,
\begin{align} 
&G_4(t_1,t_2,t_3,t_4) \nonumber\\
&=  \braket{{\cal T} \hat{O}_1\hat{O}_2\hat{O}_3\hat{O}_4}_{0,F}- \braket{{\cal T} \hat{O}_1\hat{O}_2\hat{O}_3\hat{O}_4}_{0}  \\
&= \braket{{\cal T} \hat{O}_1\hat{O}_2\hat{O}_3\hat{O}_4}_{0,F} -\sum_{\substack{3 \text{ perm.} \\ a,b,c,d\\ \in\{1,2,3,4\}}} \braket{{\cal T}\hat{O}_a\hat{O}_b}_{0}\braket{{\cal T}\hat{O}_c\hat{O}_d}_{0} \nonumber\;,
\end{align}
where we make use of the abbreviation $\hat{O}_i=\hat{O}(t_i)$.
The summation runs over all indistinguishable permutations.
With $\braket{\dots}_{0}$ and $\braket{\dots}$, we refer to correlators for which we assume Wick's theorem to be exactly valid. 
The index $0$ indicates that the system is considered without perturbation by the bath. 
In contrast to this, $\braket{\dots}_F$ ($\braket{\dots}_{0,F}$) describes the (un)perturbed correlators including the corrections to Wick's theorem. 
In this paper we consider corrections up to first order in  $G_4$.

With measurements on the quantum simulator we have access to $n$-time correlators $\braket{\dots}_F$ of $\hat{O}$. Measuring two- and four-time correlators,
\begin{equation} \label{eq_checkingwick_4er}
\braket{{\cal T} \hat{O}_{1}\hat{O}_{2}\hat{O}_{3}\hat{O}_{4}}_F - \!\!\!\!\!\!\!\!\!\! \sum_{\substack{3 \text{ perm.} \\ a,b,c,d\\ \in\{1,2,3,4\}}} \!\!\!\!\!\!\!\!\!\braket{{\cal T}\hat{O}_a\hat{O}_b}_F \braket{{\cal T}\hat{O}_c\hat{O}_d}_F 
= \;
\begin{tikzpicture}[anchor=base,baseline=5pt]
    \coordinate (A) at (0,0);
    \coordinate (B) at (0,0.5);
    \coordinate (C) at (0.5,0);
    \coordinate (D) at (0.5,0.5);
    \draw[line width=2.0pt] (A) -- (D);
    \draw[line width=2.0pt] (B) -- (C);
\end{tikzpicture} \;,
\end{equation}
we get access to the quantity,
\begin{align}
\begin{tikzpicture}[anchor=base,baseline=5pt]
    \coordinate (A) at (0,0);
    \coordinate (B) at (0,0.5);
    \coordinate (C) at (0.5,0);
    \coordinate (D) at (0.5,0.5);
    \draw[line width=2.0pt] (A) -- (D);
    \draw[line width=2.0pt] (B) -- (C);
\end{tikzpicture}
=&
\begin{tikzpicture}[anchor=base,baseline=5pt]
    \coordinate (A) at (0,0);
    \coordinate (B) at (0,0.5);
    \coordinate (C) at (0.5,0);
    \coordinate (D) at (0.5,0.5);
    \draw[line width=1.0pt] (A) -- (D);
    \draw[line width=1.0pt] (B) -- (C);
\end{tikzpicture}
+
\begin{tikzpicture}[anchor=base,baseline=5pt]
    \coordinate (A) at (0,0);
    \coordinate (B) at (0,0.5);
    \coordinate (C) at (0.5,0);
    \coordinate (D) at (0.5,0.5);
    \draw[line width=1.0pt] (A) -- (D);
    \draw[line width=1.0pt] (B) -- (C);
    \coordinate (L1a) at (1.1,0);
    \draw[line width=1.0pt, snake it] (C) -- (L1a);
    \fill (C) circle (2pt);
    \fill[white] (C) circle (1pt);
    \coordinate (L1b) at (1.7,0);
    \draw[line width=1.0pt] (L1a) -- (L1b);
    \fill (L1a) circle (2pt);
    \fill[white] (L1a) circle (1pt);
\end{tikzpicture}
+
\begin{tikzpicture}[anchor=base,baseline=5pt]
    \coordinate (A) at (0,0);
    \coordinate (B) at (0,0.5);
    \coordinate (C) at (0.5,0);
    \coordinate (D) at (0.5,0.5);
    \draw[line width=1.0pt] (A) -- (D);
    \draw[line width=1.0pt] (B) -- (C);
    \coordinate (L1a) at (1.1,0);
    \coordinate (L2a) at (1.1,0.5);
    \draw[line width=1.0pt, snake it] (C) -- (L1a);
    \draw[line width=1.0pt, snake it] (D) -- (L2a);
    \fill (C) circle (2pt);
    \fill[white] (C) circle (1pt);
    \fill (D) circle (2pt);
    \fill[white] (D) circle (1pt);
    \coordinate (L1b) at (1.7,0);
    \coordinate (L2b) at (1.7,0.5);
    \draw[line width=1.0pt] (L1a) -- (L1b);
    \draw[line width=1.0pt] (L2a) -- (L2b);
    \fill (L1a) circle (2pt);
    \fill[white] (L1a) circle (1pt);
    \fill (L2a) circle (2pt);
    \fill[white] (L2a) circle (1pt);
\end{tikzpicture}
\nonumber\\
&+
\begin{tikzpicture}[anchor=base,baseline=5pt]
    \coordinate (A) at (0,0);
    \coordinate (B) at (0,0.5);
    \coordinate (C) at (0.5,0);
    \coordinate (D) at (0.5,0.5);
    \draw[line width=1.0pt] (A) -- (D);
    \draw[line width=1.0pt] (B) -- (C);
    \coordinate (L1a) at (1.1,0);
    \draw[line width=1.0pt, snake it] (C) -- (L1a);
    \fill (C) circle (2pt);
    \fill[white] (C) circle (1pt);
    \coordinate (L1b) at (1.7,0);
    \draw[line width=1.0pt] (L1a) -- (L1b);
    \fill (L1a) circle (2pt);
    \fill[white] (L1a) circle (1pt);
    \coordinate (L1c) at (2.3,0);
    \draw[line width=1.0pt, snake it] (L1b) -- (L1c);
    \fill (L1b) circle (2pt);
    \fill[white] (L1b) circle (1pt);
    \coordinate (L1d) at (2.9,0);
    \draw[line width=1.0pt] (L1c) -- (L1d);
    \fill (L1c) circle (2pt);
    \fill[white] (L1c) circle (1pt);
\end{tikzpicture}
+\dots \nonumber \\
&+ \begin{tikzpicture}[anchor=base,baseline=5pt]
    \coordinate (A) at (0,0);
    \coordinate (B) at (0,0.5);
    \coordinate (C) at (0.5,0);
    \coordinate (D) at (0.5,0.5);
    \draw[line width=1.0pt] (A) -- (D);
    \draw[line width=1.0pt] (B) -- (C);
    \coordinate (L1a) at (1.1,0);
    \coordinate (L2a) at (1.1,0.5);
    \draw[line width=1.0pt, snake it] (C) -- (L1a);
    \draw[line width=1.0pt, snake it] (D) -- (L2a);
    \fill (C) circle (2pt);
    \fill[white] (C) circle (1pt);
    \fill (D) circle (2pt);
    \fill[white] (D) circle (1pt);
    \coordinate (L1b) at (1.7,0);
    \coordinate (L2b) at (1.7,0.5);
    \draw[line width=1.0pt] (L1a) -- (L1b);
    \draw[line width=1.0pt] (L2a) -- (L2b);
    \fill (L1a) circle (2pt);
    \fill[white] (L1a) circle (1pt);
    \fill (L2a) circle (2pt);
    \fill[white] (L2a) circle (1pt);
    \coordinate (L3a) at (-0.6,0);
    \coordinate (L3b) at (-1.1,0);
    \draw[line width=1.0pt, snake it] (A) -- (L3a);
    \draw[line width=1.0pt] (L3a) -- (L3b);
    \fill (A) circle (2pt);
    \fill[white] (A) circle (1pt);
    \fill (L3a) circle (2pt);
    \fill[white] (L3a) circle (1pt);
\end{tikzpicture}
+\dots \;, 
\end{align}
where the thin cross represents the correction $G_4$ and the sinuous lines stand for the bath correlation function (see table~\ref{tab_All_Correlators}). The central result here is that the correction to the perturbed two-time correlator can be expressed as
\begin{equation} \label{eq_checkingwick_2er}
\braket{{\cal T} \hat{O}_{1} \hat{O}_{2}}_F =
 \braket{{\cal T} \hat{O}_{1} \hat{O}_{2}}
+
\begin{tikzpicture}[anchor=base,baseline=5pt]
    \coordinate (A) at (0,0);
    \coordinate (B) at (0,0.5);
    \coordinate (C) at (0.5,0);
    \coordinate (D) at (0.5,0.5);
    \draw[line width=2.0pt] (A) -- (D);
    \draw[line width=2.0pt] (B) -- (C);
    \draw[line width=1.0pt,snake it] (D) to[out=-45,in=45] (C);
    \fill (C) circle (2pt);
    \fill[white] (C) circle (1pt);
    \fill (D) circle (2pt);
    \fill[white] (D) circle (1pt);
\end{tikzpicture} \;.
\end{equation}
Eqs.~(\ref{eq_checkingwick_4er}) and (\ref{eq_checkingwick_2er}) show that it is possible to estimate the deviation from Wick's theorem by measuring the two- and four-time correlators and combining the measured result with our knowledge of the bath correlator.
This allows us to check whether the assumption of Wick's theorem is justified and the result of the reconstruction is reliable.


\section{Imperfect knowledge}\label{sec_imperfect_measurement}

A fundamental prerequisite for the reconstruction of the unperturbed correlator is the knowledge of the perturbed correlator of the system $G_{SB}$ and the correlator of the bath $G_{B0}$.
In reality, we will not receive these quantities with full accuracy.
In this section, we address the question how imperfect knowledge affects the reconstruction of the ideal Green's function. 

\subsection{Bath correlator}
We consider a variation of the Green's function of the bath $G_{B0}(\omega)+\delta G_{B0}(\omega)$.
With this Green's function, we reconstruct the correlator of the simulator using Eq.~(\ref{eq_Essential_Result_most_simple}) with
\begin{equation}
\tilde{G}_{S0}(\omega)= \frac{G_{SB}(\omega)}{1+G_{B0}(\omega)G_{SB}(\omega)+\delta G_{B0}(\omega)G_{SB}(\omega)} \,.
\end{equation}
For $|\delta G_{B0}(\omega)|\ll| G^{-1}_{SB}(\omega)+ G_{B0}(\omega)|$, we find
\begin{equation} \label{eq_variation_bath}
\tilde{G}_{S0}(\omega) \approx G_{S0}(\omega) [1-G_{S0}(\omega) \delta G_{B0}(\omega)]\,.
\end{equation}
Hence, the impact of $\delta G_{B0}(\omega)$ is large at the peaks of $G_{S0}(\omega)$.
The influence of $\delta G_{B0}(\omega)$ is independent of the value of $G_{B0}(\omega)$.
This means that the quality of the reconstruction is defined by the absolute error $\delta G_{B0}(\omega)$ only. 

\subsection{Full system correlator}
For a deviation of the full system correlator $G_{SB}(\omega)+\delta G_{SB}(\omega)$, we have
\begin{equation}
\tilde{G}_{S0}(\omega)= \frac{G_{SB}(\omega)+ \delta G_{SB}(\omega)}{1+G_{B0}(\omega)G_{SB}(\omega)+G_{B0}(\omega) \delta G_{SB}(\omega)} \,.
\end{equation}
For  $|\delta G_{SB}(\omega)|\ll| G^{-1}_{B0}(\omega)+ G_{SB}(\omega)|$, we find
\begin{equation}
\tilde{G}_{S0}(\omega) \approx G_{S0}(\omega) \left( 1+ \frac{ G_{S0}(\omega)}{ G_{SB}(\omega)} \frac{\delta G_{SB}(\omega)}{ G_{SB}(\omega)} \right) \,.
\end{equation}
The ratio of $G_{S0}(\omega)$ and $G_{SB}(\omega)$ implies that the variation of the full system correlator $G_{S0}(\omega)$ is large at the peaks of this function.
In contrast to the variation of the bath correlator in Eq.~(\ref{eq_variation_bath}), the relative error $\delta G_{SB}(\omega)/ G_{SB}(\omega)$ enters here.

In addition, this equation shows the limit of our reconstruction method.
Consider the limit of large coupling of the bath to the system.
Eq.~(\ref{eq_Essential_Result_most_simple}) is still valid,
but a reconstruction is no longer possible if the bath widens the peaks of $G_{S0}(\omega)$ significantly.
In this case $ G_{S0}(\omega)/ G_{SB}(\omega)\gg 1$ at the peaks.
Therefore, even a small relative error in the measurement of $G_{SB}(\omega)$ makes the reconstruction of $G_{S0}(\omega)$ practically impossible.


\section{Full model and discussion}\label{sec_Full_model_and_discussion}

\subsection{Extended Model}

In this section, we extend the model to a more general scenario and discuss the derivation of our results in detail.
To make our model more realistic, we consider multiple baths.
In practice, a system consisting of $N$ coupled qubits or resonators arranged in a certain two-dimensional geometry does not couple to a single bath. 
Instead, we consider a system with multiple independent baths $H_B=\sum_{i=1}^N H_{B}^i$ with $[H_{B}^i, H_{B}^j]=0$ and a similarly adjusted coupling term. 
The full Hamiltonian can now be written in the form
\begin{eqnarray}
 H=H_S+H_C+\sum_{i=1}^N H_{B}^i \,.
\end{eqnarray}
The coupling $H_C$ between the system and the additional degrees of freedom contained in $\sum_{i}H_{B}^i$ is assumed to be of the form
\begin{eqnarray}\label{eq_Coupling_to_a_Multipartite_bath}
 H_C &=& \lambda_B\sum_{i=1}^N \hat{O}^i  \hat{X}^i\, . 
\end{eqnarray}
The system and bath variables satisfy the commutation relation $[\hat{X}^i,H_S]=[\hat{O}^i,H_B]=[\hat{X}^i,\hat{X}^j]=0$.
We have now $N$ system operators $\hat{O}^i$ which couple the system to $N$ baths via the corresponding bath operators $\hat{X}^i$.
We have introduced the dimensionless constant $\lambda_B \in \{0,1\}$, which allows us to define the free and perturbed correlators in a more rigorous way
(see Table~\ref{tab_All_Correlators}). 

To perform the reconstruction of the unperturbed Green's function of the system, we need to characterize the properties of the baths independently of the system~\cite{Wilhelm}.
This assumption is justified, for example, for a large network of superconducting flux qubits coupled in a two-dimensional (2D) structure to simulate a spin system. 
Such systems have been realized with up to 1000 qubits~\cite{D_Wave500Qubits,D_Wave1000_qubits}.
The ideal Hamiltonian in this case would be, e.g.,
$H_S=\frac{1}{2}\sum_i h_i \sigma_x^i+\sum_{ij} J_{ij}\sigma_z^i\sigma_z^j $. Here $h_i$ and $J_{ij}$ are adjustable parameters which define the model under investigation
and $\sigma_k^i$ are the Pauli matrices acting on qubit $i$. 
Under the assumption that the effect of the noise on a single qubit is almost Markovian it is possible to characterize the noise spectral density of the decoupled qubits using the method described in \cite{measure_bath}.
The qubits are coupled to individual baths, whose bath correlators  $\langle \hat{X}^i(t) \hat{X}^i(0)\rangle_0$ are known relatively well, as estimated in Ref.~[\onlinecite{D_WaveNoise}].
From a multitude of similar experiments we know that the system operator that couples to the bath corresponds to $\hat{O}^i=\sigma_z^i$. 
Thus, for such a quantum simulator the characterization of the bath correlator is possible independently of the properties of the simulator.
Furthermore, the applicability of Wick's theorem has been studied broadly~\cite{Wick_1,Wick_2,Wick_3} in context of spin systems.
Devices such as large networks of superconducting flux qubits coupled in a 2D structure
can also be tuned into alternative regimes, e.g., into a weakly nonlinear regime where proposals exist on how to use such devices for the simulation of vibronic transitions~\cite{Vibronic}.
In this limit, the application of Wick's theorem would also be more straightforward.

\subsection{The full Green's function}

In Eq.~(\ref{eq_Ideal_Correlator}) we introduced the Green's function of the system without coupling to external degrees of freedom.
In this section we consider the Green's function ${\bf G}_{SB}$ of the system coupled to its bath in matrix form with the elements
\begin{equation} \label{eq_GSB_def}
G^{ij}_{SB}(t) = -i\braket{{\cal T}\hat{O}^i(t)\hat{O}^j(0)} \,,
\end{equation}
where $\braket{\dots}$ is an expectation value of the ground state of the full system.
Using the standard technique for Green's functions at zero temperature, we expand $G^{ij}_{SB}(t)$ in orders of $H_C$.
Therefore, the zeroth-order Hamiltonian is given by $H_0=H_S+\sum_i H_{B}^i$.
We define the time evolution
\begin{equation}
 S_{\lambda_B}(t)=e^{-i H t} \,,
\end{equation}
and transform all operators $\hat{A}$ into the appropriate picture using the definition
\begin{equation}
 \hat{A}(t)=S_{\lambda_B}^{-1}(t) \, \hat{A} \, S_{\lambda_B}(t)\,.
\end{equation}
For unperturbed correlators $\braket{\dots}_0$ this transformation with $S_{\lambda_B =0}(t)~=~e^{-i H_0 t}$ defines operators in the interaction picture, while $\lambda_B=1$ denotes the full time evolution in the Heisenberg picture for the perturbed correlators $\braket{\dots}$.
The full Green's function can be written in the form
\begin{equation}
G^{ij}_{SB}(t) = -i \frac{\braket{{\cal T}S(\infty)\hat{O}^i(t)\hat{O}^j(0)}_0}{\braket{{\cal T}S(\infty)}_0} \;,
\end{equation}
with the time evolution operator
\begin{equation}
S(\infty) = {\cal T} e^{-i \int_{-\infty}^{\infty}\mathrm{d}t \, H_C(t)} \,,
\end{equation}
where we use the coupling Hamiltonian in the interaction picture.
We introduce the Fourier transform of the Green's function
\begin{equation}
G^{ij}_X(\omega) = \int\limits_{-\infty}^{\infty} \mathrm{d}t \, e^{i \omega t} G^{ij}_X(t) \,.
\end{equation}

\subsection{Diagrammatic expansion} \label{sec_diagrammatic_expansion}

\begin{table*}[t]
 \begin{tabular}{|c|c|c|p{0.5\textwidth}|}\hline
     Green's function & Matrix form & Diagram & Definition \\ \hline
  $G_{SB}^{ij}(t) = -i \langle{\cal T} \hat{O^i}(t)\hat{O^j}(0)\rangle $  & $[{\bf G}_{SB}]_{ij}=G_{SB}^{ij}$ &
    \begin{tikzpicture}[anchor=base,baseline=8pt]
    \coordinate (A) at (0,0.44);
    \coordinate (B) at (1.5,0.44);
     \coordinate (C) at (0,0.36);
    \coordinate (D) at (1.5,0.36);
    \draw[line width=1.0pt] (A) -- (B);
    \draw[line width=1.0pt] (C) -- (D);
    \end{tikzpicture}                                                & full correlator of the system operators, including the effects of the bath ($\lambda_B=1$) \\ \hline
       $G_{S0}^{ij}(t) = -i\langle{\cal T} \hat{O^i}(t)\hat{O^j}(0)\rangle_{0} $  & $[{\bf G}_{S0}]_{ij}=G_{S0}^{ij}$ &
    \begin{tikzpicture}[anchor=base,baseline=8pt]
    \coordinate (A) at (0,0.4);
    \coordinate (B) at (1.5,0.4);
     \coordinate (C) at (0,0.32);
    \coordinate (D) at (1.5,0.32);
    \draw[line width=1.0pt] (A) -- (B);
    \end{tikzpicture}                                                & free correlator of the system operators, without the effects of the bath ($\lambda_B=0$) \\ \hline
 $G_{B0}^{ij}(t) = -i\langle{\cal T} \hat{X}^i(t)\hat{X}^j(0)\rangle_0  $  & $[{\bf G}_{B0}]_{ij}=G_{B0}^{ij}$ &
  \begin{tikzpicture}[anchor=base,baseline=8pt]
    \coordinate (A) at (0,0.4);
    \coordinate (B) at (1.5,0.4);
    \draw[line width=1.0pt,snake it] (A) -- (B);
        \end{tikzpicture}                                            & free correlator of the bath, without the effects of the system ($\lambda_B=0$) \\ \hline
     \end{tabular}
 \caption{Summary of all relevant correlators and their diagrammatic representation.}
 \label{tab_All_Correlators}
     \end{table*}

        \begin{table}[b] 
 \begin{tabular}{|c|c|p{0.25\textwidth}|}\hline
         Interaction    & Diagram   & Definition    \\  \hline
            $ \sum_{i=1}^N \hat{O}_i \hat{X}^i  $  &  
  \begin{tikzpicture}[anchor=base,baseline=8pt]
     \coordinate (A) at (0,0.4);
    \coordinate (B) at (0.75,0.4);
     \coordinate (C) at (0.75,0.4);
    \coordinate (D) at (1.5,0.4);
    \draw[line width=1.0pt] (A) -- (B);
    \draw[line width=1.0pt, snake it] (C) -- (D);
     \fill (C) circle (2pt);
    \fill[white] (C) circle (1pt);
        \end{tikzpicture}                                            & Interaction between bath and system. \\ \hline
   \end{tabular}
        \caption{Each circle represents a term of the expansion in $H_C$.}\label{tab_All_Interactions}
   \end{table}

We show now the diagrammatic expansion that leads to expressions such as Eq.~(\ref{eq_Full_Equation_of_Correlator_Simplest_case}) if Wick's theorem is valid for the coupling operators.
All relevant correlators and their diagrammatic representations are shown in Table~\ref{tab_All_Correlators} and the interaction term is shown in Table~\ref{tab_All_Interactions}.

Using an expansion of $S(\infty)$ in $H_C$,  we can directly show the connection between the Green's function of the simulator perturbed by a bath $G_{SB}^{ij}$ and the unperturbed ideal Green's functions,
\begin{eqnarray}
 \begin{tikzpicture}[anchor=base,baseline=8pt]
    \coordinate (A) at (0,0.44);
    \coordinate (B) at (0.7,0.44);
     \coordinate (C) at (0,0.36);
    \coordinate (D) at (0.7,0.36);
    \draw[line width=1.0pt] (A) -- (B);
    \draw[line width=1.0pt] (C) -- (D);
    \end{tikzpicture}             
    \!\!&=&\!\!
    \begin{tikzpicture}[anchor=base,baseline=8pt]
    \coordinate (A) at (0,0.4);
    \coordinate (B) at (0.6,0.4);
    \draw[line width=1.0pt] (A) -- (B);
    \end{tikzpicture}    
    + \begin{tikzpicture}[anchor=base,baseline=8pt]
    \coordinate (A) at (0,0.4);
    \coordinate (B) at (0.6,0.4);
    \coordinate (C) at (1.2,0.4);
    \coordinate (D) at (1.8,0.4);
    \draw[line width=1.0pt] (A) -- (B);
    \draw[line width=1.0pt, snake it] (B) -- (C);
    \draw[line width=1.0pt] (C) -- (D);
       \fill (B) circle (2pt);
    \fill[white] (B) circle (1pt);
      \fill (C) circle (2pt);
    \fill[white] (C) circle (1pt);
    \end{tikzpicture}    
    +
     \begin{tikzpicture}[anchor=base,baseline=8pt]
    \coordinate (A) at (0,0.4);
    \coordinate (B) at (0.6,0.4);
    \coordinate (C) at (1.2,0.4);
    \coordinate (D) at (1.8,0.4);
    \coordinate (E) at (2.4,0.4);
    \coordinate (F) at (3,0.4);
    \draw[line width=1.0pt] (A) -- (B);
    \draw[line width=1.0pt, snake it] (B) -- (C);
    \draw[line width=1.0pt] (C) -- (D);
    \draw[line width=1.0pt, snake it] (D) -- (E);
    \draw[line width=1.0pt] (E) -- (F);
        \fill (B) circle (2pt);
    \fill[white] (B) circle (1pt);
      \fill (C) circle (2pt);
    \fill[white] (C) circle (1pt);
       \fill (D) circle (2pt);
    \fill[white] (D) circle (1pt);
      \fill (E) circle (2pt);
    \fill[white] (E) circle (1pt);
    \end{tikzpicture}  +\ldots \nonumber\\
    &=& \!\!
    \begin{tikzpicture}[anchor=base,baseline=8pt]
    \coordinate (A) at (0,0.4);
    \coordinate (B) at (0.6,0.4);
    \draw[line width=1.0pt] (A) -- (B);
    \end{tikzpicture}    
    + \begin{tikzpicture}[anchor=base,baseline=8pt]
    \coordinate (A) at (0,0.4);
    \coordinate (B) at (0.6,0.4);
    \coordinate (C) at (1.2,0.4);
      \draw[line width=1.0pt] (A) -- (B);
    \draw[line width=1.0pt, snake it] (B) -- (C);
     \fill (B) circle (2pt);
    \fill[white] (B) circle (1pt);
      \fill (C) circle (2pt);
    \fill[white] (C) circle (1pt);
    \end{tikzpicture} 
    (
     \begin{tikzpicture}[anchor=base,baseline=8pt]
    \coordinate (A) at (0,0.4);
    \coordinate (B) at (0.6,0.4);
    \draw[line width=1.0pt] (A) -- (B);
    \end{tikzpicture}    
    + \begin{tikzpicture}[anchor=base,baseline=8pt]
    \coordinate (A) at (0,0.4);
    \coordinate (B) at (0.6,0.4);
    \coordinate (C) at (1.2,0.4);
    \coordinate (D) at (1.8,0.4);
    \draw[line width=1.0pt] (A) -- (B);
    \draw[line width=1.0pt, snake it] (B) -- (C);
    \draw[line width=1.0pt] (C) -- (D);
    \fill (B) circle (2pt);
    \fill[white] (B) circle (1pt);
    \fill (C) circle (2pt);
    \fill[white] (C) circle (1pt);
    \end{tikzpicture}    
     \ldots ) 
     \nonumber\\
     &=& \!\! \begin{tikzpicture}[anchor=base,baseline=8pt]
    \coordinate (A) at (0,0.4);
    \coordinate (B) at (0.6,0.4);
    \draw[line width=1.0pt] (A) -- (B);
    \end{tikzpicture}    
    + \begin{tikzpicture}[anchor=base,baseline=8pt]
    \coordinate (A) at (0,0.4);
    \coordinate (B) at (0.6,0.4);
    \coordinate (C) at (1.2,0.4);
     \coordinate (A1) at (1.2,0.44);
    \coordinate (B1) at (2,0.44);
     \coordinate (C1) at (1.2,0.36);
    \coordinate (D1) at (2,0.36);
    \draw[line width=1.0pt] (A1) -- (B1);
    \draw[line width=1.0pt] (C1) -- (D1);
    \draw[line width=1.0pt] (A) -- (B);
    \draw[line width=1.0pt, snake it] (B) -- (C);
     \fill (B) circle (2pt);
    \fill[white] (B) circle (1pt);
     \fill (C) circle (2pt);
    \fill[white] (C) circle (1pt);
    \end{tikzpicture} \;.
\end{eqnarray}
Here all disconnected diagrams are canceled by the vacuum diagrams in $\braket{ {\cal T} S(\infty)}_0$ (see Appendix~\ref{appendix_disconnected_diagrams}).
Therefore we can write the Dyson equation in matrix form as
\begin{equation}\label{eq_Multipartite_Bath_only_Solution}
 {\bf G}_{SB}(\omega)={\bf G}_{S0}(\omega)+{\bf G}_{S0}(\omega){\bf G}_{B0}(\omega){\bf G}_{SB}(\omega) \,.
\end{equation}
If all Green's functions $G_{SB}^{ij}(\omega)$ and $G_{B0}^{ij}(\omega)$ are known, this equation can be solved for ${\bf G}_{S0}$:
\begin{equation}
 {\bf G}_{S0}(\omega)={\bf G}_{SB}(\omega)\left[1+{\bf G}_{B0}(\omega){\bf G}_{SB}(\omega)\right]^{-1} \label{eq_GSO} \,.
\end{equation}
If we reduce the system to a single-bath situation, this result transforms to Eq.~(\ref{eq_Essential_Result_most_simple}). 
It connects the ideal correlator in Eq.~(\ref{eq_Ideal_Correlator}) to quantities which can be readily measured.

\subsection{Extension to finite temperatures} \label{sec_finite_temperatures}
The diagrammatic expansion in Sec.~\ref{sec_diagrammatic_expansion} can also be applied to the Matsubara Green's functions $\mathcal{G}_{M,X}$, which are connected to the retarded Green's functions for finite temperatures $\mathcal{G}^R_{X}$.
This is a way to extend this method to systems in thermal equilibrium.
The analog of Eq.~(\ref{eq_Essential_Result_most_simple}) for finite temperatures is given by
\begin{equation}\label{eq_Essential_Result_most_simple_Matsubara} 
 \mathcal{G}^R_{S0}(i\omega_n) = \frac{\mathcal{G}^R_{SB}(i\omega_n)}{1+\mathcal{G}^R_{B0}(i\omega_n)\mathcal{G}^R_{SB}(i\omega_n)}\,.
\end{equation}
Below we introduce the Matsubara Green's functions and explain the connection to the spectral function.

\subsubsection{Expansion in imaginary time}
As we consider the whole system to be in thermal equilibrium, it is reasonable to use the standard Matsubara Green's function method.
Therefore, we define the imaginary time $\tau=i t$ where we require $0<\tau<\beta$. 
The Matsubara Green's function equivalent to Eq.~(\ref{eq_GSB_def}) is
\begin{equation}
\mathcal{G}^{ij}_{M,SB}(\tau)= -\langle{\cal T} \hat{O}^i(\tau)\hat{O}^j(0)\rangle \,,
\end{equation}
where ${\cal T}$ is the time-ordering operator for $\tau$. 
In the case of finite temperatures, $\braket{\dots}$ refers to the equilibrium expectation value $\mathrm{Tr}(\frac{1}{Z}e^{-\beta H}\dots)$, with $Z=\mathrm{Tr}(e^{-\beta H})$. 
The time evolution in imaginary time is given by 
\begin{equation}
 U_{\lambda_B}(\tau)=e^{-H\tau} \,.
\end{equation}
We transform all operators $\hat{A}$ into the appropriate picture in imaginary time using the definition
\begin{equation}
 \hat{A}(\tau)=U_{\lambda_B}^{-1}(\tau) \, \hat{A} \, U_{\lambda_B}(\tau)\,.
\end{equation}
The full correlator can be written in the form
\begin{equation}\label{eq_Full_correlator_In_expansion_from_Matsubara}
\mathcal{G}_{M,SB}(\tau)=-\frac{\langle {\cal T} U(\beta) \hat{O}^i(\tau) \hat{O}^j(0) \rangle_0}
               {\langle {\cal T} U(\beta)\rangle_0} \,,
\end{equation}
with evolution operator
\begin{equation}
U(\tau)={\cal T} e^{-\int_0^{\tau}d\tau' H_{C,I}(\tau')} \,.
\end{equation}
As for zero temperature, all disconnected diagrams are canceled by the factor $\braket{ {\cal T} U(\beta)}_0$, the so-called vacuum diagrams.

The correlator in imaginary time is periodic in $\tau$ with period $\beta$. 
It is convenient to transform it to frequency space using the discrete Fourier transform,
\begin{equation}
\mathcal{G}^{ij}_{M,X}(\tau)=\frac{1}{\beta}\sum_n \mathcal{G}^{ij}_{M,X}(\omega_n)e^{-i\omega_n \tau}
\end{equation}
with the Matsubara frequencies $\omega_n=2\pi n/\beta$.

\subsubsection{Connecting a real time correlator to the Matsubara Green's function}\label{subsec_connecting_real_to_imaginary}
Now we discuss the connection of the Matsubara Green's function to measurable quantities such as the spectral function or correlators. As an example we focus on the Green's function of the bath. 

We define the correlation function
\begin{equation}
C^{i}(t)= \left(\braket{\hat{X}^i(t)\hat{X}^i(0)}_0-\braket{\hat{X}^i(0)\hat{X}^i(t)}_0 \right) \theta(t)\,.
\end{equation}
The eigenstates of the bath are given by $|n\rangle$, with $H_B|n\rangle=E_n|n\rangle$. This allows us to rewrite the correlator,
\begin{equation}
 C^{i}(t)=\frac{\theta(t)}{Z_B}\sum_{nm}|\langle n|\hat{X}^i|m\rangle|^2 e^{i(E_n-E_m)t}(e^{-\beta E_n} -e^{-\beta E_m}),
\end{equation}
with the partition function $Z_B={\rm Tr}(e^{-\beta H_B})$.
The real part of the Fourier transform of the correlator gives us the spectral function
\begin{eqnarray}
 A^{i}(\omega) &=& \frac{1}{\pi} {\rm Re}\left(\int_{-\infty}^{\infty} \mathrm{d}t \, e^{i\omega t} C^i(t)\right)\\
           &=&\!\!
           \frac{1}{Z_B}\sum_{nm}|\langle n|\hat{X}^i|m\rangle|^2\nonumber\\
           & & (e^{-\beta E_m} -e^{-\beta E_n}) \,\delta\!\left[\omega-(E_n\!-\!E_m)\right] \,.
           \nonumber
\end{eqnarray}
Apart from a factor $-1$, the imaginary part of the retarded Green's function $\mathcal{G}_{B0}^R(t)$ is equivalent to the correlation function $C^i(t)$, since
\begin{equation}
\mathcal{G}_{B0}^{R, ii}(t) = - i \langle [\hat{X}^i(t),\hat{X}^i(0)]\rangle_0 \theta(t) \,.  
\end{equation}
Assuming that $A^{i}(\omega)$ has been measured, the retarded Green's function $\mathcal{G}_{B0}^R(\omega)$ of the bath can be calculated using
\begin{equation} \label{eq_GBOR_S}
 \mathcal{G}_{B0}^{R, ii}(\omega)=\int_{-\infty}^{\infty} \mathrm{d}\omega_1\frac{A^i(\omega_1)}{\omega -\omega_1+i0} \,.
\end{equation}
Describing the Matsubara Green's function in terms of the spectral function shows a connection to the retarded Green's function for finite temperatures $\mathcal{G}^{R}_{P}$,
\begin{equation}
\mathcal{G}^{ij}_{M,P}(\omega_n) = \mathcal{G}^{R,ij}_{P}(i \omega_n) \;,\; \omega_n>0 \,, \label{eq_GM_equals_GR}
\end{equation}
with $P\in\{B0,S0,SB\}$.
This requires an analytic continuation of $\mathcal{G}^{R}_{P}$ in the complex plane.
Via the spectral function we can derive the Kramers-Kronig relation,
\begin{equation}
\mathcal{G}^{ij}_{P}(\omega) = {\rm Re}\mathcal{G}^{R,ij}_{P}(\omega)+ i (1+2 \bar{n}(\omega)) {\rm Im}\mathcal{G}^{R,ij}_{P}(\omega)\,,
\end{equation}
with $\bar{n}(\omega) = (e^{\beta\omega}-1)^{-1}$.
Starting from the Matsubara Green's function, we obtain information about the retarded Green's function at the points $i\omega_n$.
We would like to have the ideal Green's function $\mathcal{G}^R_{S0}$, i.e., the spectral function, for the complete real axis. 
This can be achieved by using numerical methods like the Pad\'{e} approximation approach~\cite{Pade_1,Pade_2}.
However, it should be emphasized that the numerical transformation of a Green's function at the Matsubara frequencies to the real axis 
is still a non trivial problem and an active research field~\cite{Analytical_continuatioon}.

\subsection{Model system: chain of resonators with individual baths}

In this section we give an explicit example of our method and particularly of the validity of Eq.~(\ref{eq_GSO}).
We consider a system of coupled harmonic oscillators,
\begin{equation}
H_S = \sum_{j=1}^N\left( \frac{1}{2} m\omega_r^2 q_j^2 + \frac{1}{2m} p_j^2 + \frac{m \Omega^2}{2}(q_{j+1}-q_j)^2 \right) \,,
\end{equation}
where $N$ is the number of resonators, $m$ refers to the mass, $\omega_r$ is the eigenfrequency of an uncoupled resonator, and $\Omega$ describes the coupling between neighboring oscillators. We assume periodic boundary conditions.
For a system of coupled resonators, Wick's theorem stated in Eq.~(\ref{eq_Wick_Theorem_For_O}) is clearly valid.
Here we show the validity of our previously derived results.
We validate our results for the connection between the ideal and perturbed correlators by using the quantum regression theorem (QRT)~\cite{Carmichael}.
While the system of bare coupled resonators would not make for a good quantum simulator, proposals exist for modeling the Bose-Hubbard model using coupled non linear resonators~\cite{Hartmann}.
Similarly, limiting cases from non interacting bosons to hard-core bosons have been studied in the context of analog quantum simulation~\cite{Bose_Hubbard_Cirac}.

We assume that each of the resonators is coupled to an individual bosonic bath,
\begin{align}
H_C &= \sum_j \hat{O}^j \hat{X}^j \,, \quad H_B = \sum_{j,m} \bar{\omega}_m^{(j)} b_m^{(j)\dagger} b_m^{(j)} \,, \\ \nonumber
 &\text{with} \ \hat{O}^j=q_j\,, \ \text{and} \ \hat{X}^j=\sum_m t_m^{(j)}(b_m^{(j)\dagger} + b_m^{(j)}) \;.
\end{align}
We assume the baths to be identical, i.e.,
\begin{equation}
\bar{\omega}^{(j)}_m=\bar{\omega}_m \;,\quad t_m^{(j)}=t_m\; ,
\end{equation}
but independent
\begin{equation}
\braket{\hat{X}^{j_1}(t_1)\hat{X}^{j_2}(t_2)}_0 = 0 \ \text{for} \ j_1 \neq j_2 \, .
\end{equation}
Diagonalizing the system Hamiltonian results in
\begin{equation}
H_S = \sum_k \Omega_k a_k^\dagger a_k\, , \ \text{with} \ \Omega_k = \sqrt{\left[2 \Omega \sin(k\frac{\varphi_0 }{2})\right]^2+\omega_r^2} \,,
\end{equation}
where $\varphi_0=\frac{2 \pi}{N}$.
The connection of annihilation and creation operators of system eigenstates, $a_k$ and $a_k^\dagger$, to the original operators has the form
\begin{alignat}{3}
q_j &= \sqrt{\frac{1}{2m\omega_r}} && \!\!\!( d_j^\dagger + d_j ) \,, \\
d_j &= \frac{1}{2\sqrt{N}} \sum_{k=1}^N &&\left[ e^{-ikj\varphi_0} \left( \sqrt{\frac{\omega_r}{\Omega_K}} -\sqrt{\frac{\Omega_K}{\omega_r}} \right) a_k^\dagger \right. \nonumber\\
&&&+ \left. e^{ijk\varphi_0} \left( \sqrt{\frac{\omega_r}{\Omega_K}} + \sqrt{\frac{\Omega_K}{\omega_r}} \right) a_k \right]  \, .
\end{alignat}
We consider finite temperatures. Therefore, the spectral density of the bath is given by
\begin{equation}
A^i(\omega) \approx \frac{1}{2\pi} \mathrm{sign}(\omega) J^i(|\omega|) \,,
\end{equation}
with  $J^i(\omega) = J(\omega) = 2\pi \sum_m t_m^{2} \delta(\omega-\bar{\omega}_m)$.

To compare Eq.~(\ref{eq_GSO}) to correlators calculated using a master-equation approach,
 we calculate the full Green's function $\mathcal{G}^{j_1 j_2}_{M,SB}$ using the QRT.
To this end we assume the dynamics of the full system to be approximately described by the Lindblad equation
\begin{equation}
\dot{\rho}(t) = \mathcal{L}\rho(t) \, ,
\end{equation}
with the Lindblad terms
\begin{align}
\mathcal{L}\rho =& -i [H_S,\rho] \nonumber\\
&+  \sum_{k=1}^N \frac{\Gamma_k}{2} (\bar{n}_k +1) \left( 2 a_k \rho a_k^\dagger - a_k^\dagger a_k \rho - \rho a_k^\dagger a_k \right) \nonumber\\
 &+ \sum_{k=1}^N \frac{\Gamma_k}{2} \bar{n}_k \left( 2 a_k^\dagger \rho a_k - a_k a_k^\dagger \rho - \rho a_k a_k^\dagger \right) \,, \label{eq_Lindblad_Resonators}
\end{align}
where $\bar{n}_k=(e^{\beta\Omega_k}-1)^{-1}$.
Assuming the spectral density of the bath to be smooth, we find the effective rates
\begin{equation}
\Gamma_k = \frac{1}{2m\Omega_k} J(\Omega_k) \,,
\end{equation}
where the prefactor $(2m\omega_r)^{-1}$ arises from $\hat{O}^i= \sqrt{2m\omega_r}^{-1} (d_j^{\dagger}+d_j)$ and $\frac{\omega_r}{\Omega_k}$ is a result of the transition from $d_j^{\dagger}+d_j$ to $a_k^{\dagger}+a_k$.
In accordance with the assumptions used for the Lindblad equation, Eq.~(\ref{eq_GBOR_S}) reduces to
\begin{equation} \label{eq_GBOR_J}
i\mathcal{G}_{B0}^{R, ij}(\omega)\approx\delta_{ij} \frac{1}{2}\mathrm{sign}(\omega)J^i(|\omega|) \,.
\end{equation}
For the Lindblad equation to be valid, some assumptions have to be made about the 
spectral density of the bath.

With the QRT, the Lindblad terms fulfill the following equation for an arbitrary operator $\hat{A}$ and all $k$~\cite{Carmichael}:
\begin{equation}
{\rm Tr}\left[ a_k \mathcal{L} \hat{A} \right] = -(i\Omega_k +\frac{\Gamma_k}{2}) {\rm Tr}\left[ a_k \hat{A}\right] \, .
\end{equation}
For $t>0$ we get
\begin{align}
\braket{\hat{A}(t_0) a_{k}(t + t_0)} &= e^{-i \Omega_k t} e^{- \frac{\Gamma_k}{2} t}\braket{\hat{A}(t_0) a_{k}(t_0)} \,,\\
\braket{a_{k}(t + t_0)\hat{A}(t_0)} &= e^{-i \Omega_k t} e^{- \frac{\Gamma_k}{2} t}\braket{a_{k}(t_0)\hat{A}(t_0)}\,,\\
\braket{\hat{A}(t_0) a_{k}^\dagger(t + t_0)} &= e^{+i \Omega_k t} e^{- \frac{\Gamma_k}{2} t}\braket{\hat{A}(t_0) a_{k}^\dagger(t_0)} \,,\\
\braket{a_{k}^\dagger(t + t_0)\hat{A}(t_0)} &= e^{+i \Omega_k t} e^{- \frac{\Gamma_k}{2} t}\braket{a_{k}^\dagger(t_0)\hat{A}(t_0)} \, .
\end{align}
The stationary solution of the Lindblad equation is proportional to $e^{-\beta \sum_k\Omega_k a_k^\dagger a_k}$.
Using this, we calculate the initial values for $\braket{a_{k}^{(\dagger)}(t_0)a_k^{(\dagger)}(t_0)}$ and find
\begin{align}
\braket{a_k(t_1)a_{k'}(t_2)} &= 0\,,\\
\braket{a_k^\dagger(t_1)a_{k'}(t_2)} &= \delta_{k,k'} \bar{n}_k e^{i\Omega_k (t_1-t_2)} e^{-\frac{\Gamma_k}{2} |t_1-t_2|}\,,\\\
\braket{a_k(t_1)a_{k'}^\dagger(t_2)} &= \delta_{k,k'} (\bar{n}_k + 1) e^{i\Omega_k (t_1-t_2)} e^{-\frac{\Gamma_k}{2} |t_1-t_2|}\,,\\
\braket{a_k^\dagger(t_1)a_{k'}^\dagger(t_2)} &= 0 \,.
\end{align}
A direct calculation of the free correlators results in
\begin{align}
\braket{a_k(t_1)a_{k'}(t_2)}_0 &= 0\,,\\
\braket{a_k^\dagger(t_1)a_{k'}(t_2)}_0 &= \delta_{k,k'} \bar{n}_k e^{i\Omega_k (t_1-t_2)} \,,\\\
\braket{a_k(t_1)a_{k'}^\dagger(t_2)}_0 &= \delta_{k,k'} (\bar{n}_k + 1) e^{i\Omega_k (t_1-t_2)}\,,\\
\braket{a_k^\dagger(t_1)a_{k'}^\dagger(t_2)}_0 &= 0 \, .
\end{align}
From this result we calculate the retarded Green's functions $\mathcal{G}^{R,j_1j_2}_{S0}(t)$, $\mathcal{G}^{R,j_1j_2}_{SB}(t)$ and perform the Fourier transform. With an analytic continuation and Eq.~(\ref{eq_GM_equals_GR}) we finally arrive at the Matsubara Green's functions for $\omega_n>0$:
\begin{align}
\mathcal{G}^{j_1j_2}_{M,SO}(\omega_n) =& \frac{1}{N} \sum_{k=1}^N \frac{1}{2 m \Omega_k} \nonumber\\
&\times \left[ e^{-i k(j_1-j_2)\varphi_0}\bar{n}_k -e^{i k(j_1-j_2)\varphi_0}(\bar{n}_k +1) \right] \nonumber\\
&\times \left( \frac{1}{i \omega_n +\Omega_k+ i0} -\frac{1}{i \omega_n -\Omega_k+ i0}\right) \,,\\
\mathcal{G}^{j_1j_2}_{M,SB}(\omega_n) =& \frac{1}{N} \sum_{k=1}^N \frac{1}{2 m \Omega_k} \nonumber\\
&\times \left[ e^{-i k(j_1-j_2)\varphi_0}\bar{n}_k -e^{i k(j_1-j_2)\varphi_0}(\bar{n}_k +1) \right] \nonumber\\
&\times \left( \frac{1}{i \omega_n +\Omega_k+ i\frac{\Gamma_k}{2}} -\frac{1}{i \omega_n -\Omega_k+ i\frac{\Gamma_k}{2}}\right) \,.
\end{align}
To calculate the bath Green's function using Eq.~(\ref{eq_GSO}) we introduce the transformation
\begin{align}
\mathcal{G}^{k}_{M,S0}(\omega_n) =& \sum_{j_1,j_2} \mathcal{G}^{j_1j_2}_{M,SO} e^{ik(j_1-j_2)k\varphi_0}\nonumber\\
 =&\frac{N}{2m\Omega_k} \left( \frac{1}{i \omega_n -\Omega_k+ i0} -\frac{1}{i \omega_n +\Omega_k+ i0} \right) \,, \\
\mathcal{G}^{k}_{M,SB}(\omega_n) =& \sum_{j_1,j_2} \mathcal{G}^{j_1j_2}_{M,SB} e^{ik(j_1-j_2)k\varphi_0}\nonumber\\
 =&\frac{N}{2m\Omega_k} \left( \frac{1}{i \omega_n -\Omega_k+ i\frac{\Gamma_k}{2}} -\frac{1}{i \omega_n +\Omega_k+ i\frac{\Gamma_k}{2}} \right) \,.
\end{align}
With this Eq.~(\ref{eq_GSO}) results in
\begin{align}
\mathcal{G}^{k}_{M,SB}(\omega_n) = \mathcal{G}^{k}_{M,S0}(\omega_n) +&  \mathcal{G}^{k}_{M,S0}(\omega_n) \mathcal{G}^{k}_{M,SB}(\omega_n) \nonumber\\ 
&\times\sum_j \mathcal{G}^{jj}_{M,BO}(\omega_n) \frac{1}{N^2} \,.
\end{align}
In the Lindblad equation we take into account the spectral density of the bath at $\Omega_k$. Since the bath Green's function depends on the spectral density of the bath, the relation is true for $\omega_n\approx\Omega_k$. Using the assumption of identical and independent baths we arrive at
\begin{equation}
\mathcal{G}^{j_1j_2}_{M,BO}(\omega_n\approx \Omega_k) \approx \delta_{j_1,j_2}  m \Gamma_k\left( \Omega_k +\frac{\Gamma_k}{4} \right) \, .
\end{equation}
In the limit of small coupling to the bath $\Gamma_k \ll \Omega_k$, we are left with
\begin{equation}
\mathcal{G}^{j_1j_2}_{M,BO}(\omega_n\approx\Omega_k)\approx \delta_{j_1,j_2} \frac{1}{2} J(\Omega_k) \,,
\end{equation}
From a comparison to Eq.~(\ref{eq_GBOR_J}), we conclude that Eq.~(\ref{eq_GSO}) holds for this example. For an Ohmic spectral density and with $\Omega_k \rightarrow i\omega$ the Matsubara Green's function of the bath coincides with Eq.~(\ref{eq_GBOR_J}).

\vspace*{1em}


\section{Conclusions}
The main result we presented in this paper is twofold.
On the one hand, we introduced a method that can be used to reconstruct certain unperturbed (ideal) Green's functions from the perturbed ones, measured by a quantum simulator coupled to additional degrees of freedom.
To achieve this, we assumed that any $n$-time correlator of the coupling operator of the ideal system can be written as a product of two-time correlators. This is known as Wick's theorem. On the other hand, we explained how to verify this assumption by a measurement.
Furthermore, we assumed good knowledge of the bath correlators to perform the reconstruction.
In particular, we presumed that these correlators are measured independently when not coupled to the ideal system.
We also clarified how imperfect measurements of the bath and of the full correlator affect the reconstruction. 
For example, in the case of strong coupling to the bath our result is still valid, but the reconstruction fails even in the presence of small noise during the measurement.

Presently, the applicability of analog quantum simulation is severely restricted, since the influence of sources of errors is not well understood. 
The approach presented in this paper leads the way to quantify and even correct errors in quantum simulation.
Since the reconstruction method is based on classical postprocessing, this method helps to make the results of quantum simulation reliable without adding an overhead to the quantum system.
Therefore, the promising potential of quantum simulation to yield interesting results even using small quantum systems remains.

\begin{acknowledgements}
The authors thank Daniel Mendler, Christian Karlewski and Gerd Sch\"on for enlightening discussions. I.S. acknowledges financial support by Friedrich-Ebert-Stiftung. L.T. was supported by the National Science Foundation (USA) under Award Number DMR-0956064, the UC Multicampus-National Lab Collaborative Research and Training under Award No. LFR-17-477237, and the UC Merced Faculty Research Grants 2017.
\end{acknowledgements}


\begin{widetext}
\appendix
\section{Disconnected Diagrams}
\label{appendix_disconnected_diagrams}
In this section, we explain how the so-called vacuum diagrams $\braket{{\cal T} S(\infty)}_0$  cancel the disconnected diagrams in the free two-time correlator $\braket{ {\cal T} S(\infty) \hat{O}_I(t) \hat{O}_I(0)}_0$.
To shorten the equations we use $\hat{A}_i$ as an abbreviation for $\hat{A}(t_i)$.
For simplicity we base our discussion on a coupling Hamiltonian of the form $H_C=\hat{O}\hat{X}$.
It is straight forward to extend this calculations on the full model described in Sec.~(\ref{sec_Full_model_and_discussion}).
The vacuum diagrams are given by
\begin{equation} 
\braket{{\cal T} S(\infty)}_0 = \sum_n \frac{1}{n!} (-i)^n \int\limits_{-\infty}^\infty \mathrm{d}t_1 \dots \int\limits_{-\infty}^\infty \mathrm{d}t_n  \braket{{\cal T} \hat{O}_1 \dots \hat{O}_n}_0 \braket{{\cal T} \hat{X}_1 \dots \hat{X}_n}_0  = \sum_n V_n  \,,
\end{equation}
where we assume
\begin{equation} 
\braket{\hat{O}_I(t)}_0 = 0 \,, \quad \braket{\hat{X}_{I}(t)}_0 = 0  \,,
\end{equation}
so that terms with $n$ being an odd number are zero.
We have introduced $V_n$, the vacuum diagrams of order $n$.
Now we elaborate the connection between the free correlator and the vacuum diagrams.
The free two-time correlator is given by
\begin{equation}
\braket{{\cal T}S(\infty) \hat{O}_a \hat{O}_b}_0 = \sum_n \frac{1}{n!} (-i)^n \int\limits_{-\infty}^\infty \mathrm{d}t_1 \dots \int\limits_{-\infty}^\infty \mathrm{d}t_n  \braket{{\cal T} \hat{O}_a \hat{O}_b\hat{O}_1 \dots \hat{O}_n}_0  \braket{{\cal T} \hat{X}_1 \dots \hat{X}_n}_0 \,.
\end{equation}
From this we apply Wick's theorem and take out the two-time correlators which form a connected diagram and recombine the surplus correlators in a higher-order correlator.
There are $\frac{n!}{(n-m)!}$ possibilities to choose $m$ vertices out of $n$.
Therefore, a connected diagram with $m$ vertices occurs $\frac{n!}{(n-m)!}$ times
\begin{eqnarray}
\braket{{\cal T}S(\infty) \hat{O}_a \hat{O}_b}_0 = \sum_n \frac{1}{n!} (-i)^n \!\!\!\int\limits_{-\infty}^\infty \mathrm{d}t_1 \dots\! \int\limits_{-\infty}^\infty \mathrm{d}t_n \sum_m^n &&  \!\!\!\!  \braket{{\cal T}\hat{O}_a\hat{O}_1}_0 \braket{{\cal T}\hat{X}_1\hat{X}_2}_0 \braket{{\cal T}\hat{O}_2\hat{O}_3}_0 \dots  \braket{{\cal T}\hat{X}_{m-1}\hat{X}_m}_0 \braket{{\cal T}\hat{O}_m\hat{O}_b}_0  \nonumber\\
&&\cdot {\textstyle \frac{n!}{(n-m)!}}  \braket{{\cal T} \hat{O}_{m+1} \dots \hat{O}_n}_0  \braket{{\cal T} \hat{X}_{m+1} \dots \hat{X}_n}_0 \,.
\end{eqnarray}
By resorting the factors we can identify the vacuum diagrams of order $n-m$,
\begin{eqnarray}
\braket{{\cal T}S(\infty) \hat{O}_a \hat{O}_b}_0=\sum_n \sum_m^n &
\underbrace{
(-i)^m \!\!\!\int\limits_{-\infty}^\infty \mathrm{d}t_1 \dots\! \int\limits_{-\infty}^\infty \mathrm{d}t_m  \braket{{\cal T}\hat{O}_a\hat{O}_1}_0  \braket{{\cal T}\hat{X}_1\hat{X}_2}_0 \braket{{\cal T}\hat{O}_2\hat{O}_3}_0 \dots  \braket{{\cal T}\hat{X}_{m-1}\hat{X}_m}_0 \braket{{\cal T}\hat{O}_m\hat{O}_b}_0
}_{=C^{a,b}_m} \nonumber\\
&\cdot \overbrace{
{\textstyle\frac{1}{(n-m)!}} (-i)^{n-m}  \int\limits_{-\infty}^\infty \mathrm{d}t_{m+1} \dots\! \int\limits_{-\infty}^\infty \mathrm{d}t_n \braket{{\cal T} \hat{O}_{m+1} \dots \hat{O}_n}_0  \braket{{\cal T} \hat{X}_{m+1} \dots \hat{X}_n}_0
}^{=V_{n-m}} \,,
\end{eqnarray}
and find the connected diagrams of order $m$, which we will call $C^{a,b}_m$, with  $C^{a,b}_0=\braket{{\cal T} \hat{O}_a \hat{O}_b}_0$.
One can factor out $\braket{{\cal T} S(\infty)}_0$ by using the Cauchy product formula
\begin{equation}
\braket{{\cal T}S(\infty) \hat{O}_a \hat{O}_b}_0=\sum_n^\infty \sum_m^n C^{a,b}_m V_{n-m} = \sum_m^\infty C^{a,b}_m\sum_n^\infty V_n = \braket{{\cal T} S(\infty)}_0 \sum_m^\infty C^{a,b}_m \,.
\end{equation}
This means, that the vacuum diagrams cancel all disconnected diagrams, i.e.,
\begin{equation}
\frac{\braket{{\cal T}S(\infty) \hat{O}_a \hat{O}_b}_0}{\braket{{\cal T} S(\infty)}_0} =  
    \begin{tikzpicture}[anchor=base,baseline=8pt]
    \coordinate (A) at (0,0.4);
    \coordinate (B) at (0.6,0.4);
    \draw[line width=1.0pt] (A) -- (B);
    \end{tikzpicture}    
    + \begin{tikzpicture}[anchor=base,baseline=8pt]
    \coordinate (A) at (0,0.4);
    \coordinate (B) at (0.6,0.4);
    \coordinate (C) at (1.2,0.4);
    \coordinate (D) at (1.8,0.4);
    \draw[line width=1.0pt] (A) -- (B);
    \draw[line width=1.0pt, snake it] (B) -- (C);
    \draw[line width=1.0pt] (C) -- (D);
       \fill (B) circle (2pt);
    \fill[white] (B) circle (1pt);
      \fill (C) circle (2pt);
    \fill[white] (C) circle (1pt);
    \end{tikzpicture}
    + \begin{tikzpicture}[anchor=base,baseline=8pt]
    \coordinate (A) at (0,0.4);
    \coordinate (B) at (0.6,0.4);
    \coordinate (C) at (1.2,0.4);
    \coordinate (D) at (1.8,0.4);
    \coordinate (E) at (2.4,0.4);
    \coordinate (F) at (3.0,0.4);
    \draw[line width=1.0pt] (A) -- (B);
    \draw[line width=1.0pt, snake it] (B) -- (C);
    \draw[line width=1.0pt] (C) -- (D);
    \draw[line width=1.0pt, snake it] (D) -- (E);
    \draw[line width=1.0pt] (E) -- (F);
       \fill (B) circle (2pt);
    \fill[white] (B) circle (1pt);
      \fill (C) circle (2pt);
    \fill[white] (C) circle (1pt);
       \fill (D) circle (2pt);
    \fill[white] (D) circle (1pt);
       \fill (E) circle (2pt);
    \fill[white] (E) circle (1pt);
    \end{tikzpicture}
    + \dots \,.
\end{equation}

\section{Four-time correlator} \label{app_4-time_correlator}
In this section, we consider a system where Wick's theorem is not exactly valid. The goal is to derive Eqs.~(\ref{eq_checkingwick_4er}) and (\ref{eq_checkingwick_2er}), in order to quantify the deviation from Wick's theorem.  We define the lowest order correction to Wick's theorem as $G_4(t_1,t_2,t_3,t_4)$,
\begin{equation}
G_4(t_1,t_2,t_3,t_4) =  \braket{{\cal T} \hat{O}_1\hat{O}_2\hat{O}_3\hat{O}_4}_{0,F}- \braket{{\cal T} \hat{O}_1\hat{O}_2\hat{O}_3\hat{O}_4}_{0}= \braket{{\cal T} \hat{O}_1\hat{O}_2\hat{O}_3\hat{O}_4}_{0,F} -\sum_{\substack{3 \text{ perm.} \\ a,b,c,d}} \braket{{\cal T}\hat{O}_a\hat{O}_b}_{0}\braket{{\cal T}\hat{O}_c\hat{O}_d}_{0} \; ,
\end{equation}
where the summation runs over all three indistinguishable permutations.
With $\braket{\dots}$ ($\braket{\dots}_{0}$) we refer to (un)perturbed correlators for which we assume Wick's theorem to be exactly valid. In contrast to this,  $\braket{\dots}_F$ ($\braket{\dots}_{0,F}$) describe the (un)perturbed correlators including the corrections to Wick's theorem. In this paper we only consider the lowest-order correction to Wick's theorem ($G_4$).
All higher-order corrections are neglected.
To shorten the equations we use the abbreviation  $G_4(1,2,3,4)=G_4(t_1,t_2,t_3,t_4)$.
An $n$-time correlator is then given by
\begin{equation}
\braket{{\cal T} \hat{O}_1 \dots \hat{O}_n}_{0,F} = \braket{{\cal T} \hat{O}_1 \dots \hat{O}_n}_{0} + \sum_{\substack{\text{perm.}\\ \alpha,\beta,\gamma,\delta}} G(\alpha,\beta,\gamma,\delta) \braket{{\cal T} \, \prod\limits_{\substack{k \in \{1,\dots n\}\backslash\{\alpha,\beta,\gamma,\delta\}}} \hat{O}_k}\!{}_{0} \label{eq_correctiontowick} \,.
\end{equation}

At first we show for the four-time correlator that if Wick's theorem is valid for the unperturbed correlator, it is also valid for the perturbed one. We start with
\begin{align}
\braket{{\cal T} \hat{O}_{\text{I}}\hat{O}_{\text{II}}\hat{O}_{\text{III}}\hat{O}_{\text{IV}}} &= \frac{\braket{{\cal T}S(\infty) \hat{O}_{\text{I}} \hat{O}_{\text{II}}\hat{O}_{\text{III}}\hat{O}_{\text{IV}}}_{0}}{\braket{{\cal T} S(\infty)}_{0}} \\
&= \sum_n \frac{(-i)^n}{n!} \int_{-\infty}^\infty \mathrm{d}t_1 \dots\! \int\limits_{-\infty}^\infty \mathrm{d}t_n \frac{1}{\braket{{\cal T} S(\infty)}_{0}}\braket{{\cal T}\hat{O}_{\text{I}} \hat{O}_{\text{II}}\hat{O}_{\text{III}}\hat{O}_{\text{IV}}\hat{O}_1 \dots \hat{O}_n}_{0} \braket{{\cal T}\hat{X}_1 \dots \hat{X}_n}_{0} \;.
\end{align}
We focus on a coupling Hamiltonian of the form $H_C=\hat{O} \hat{X}$. 
We proceed as in the above section and identify connected diagrams $C^{a,b}_m$ with $m$ vertices. Such diagrams occur $\frac{n!}{(n-m)!}$ times. There are six indistinguishable possibilities to choose $a$ and $b$. Out of the remaining $n-m$ operators we choose a connected diagram $C^{c,d}_k$ with $k$ vertices. This occurs $\frac{(n-m)!}{(n-m-k)!}$ times. As, for example, $C^{\text{I},\text{II}}_m$ and $C^{\text{I},\text{II}}_k$ for $m=k$ are indistinguishable, we have in fact three indistinguishable permutations to take into account:
\begin{align}
&\braket{{\cal T} \hat{O}_{\text{I}}\hat{O}_{\text{II}}\hat{O}_{\text{III}}\hat{O}_{\text{IV}}} \nonumber\\ 
&=\sum_n \frac{(-i)^n}{n!} \!\! \int\limits_{-\infty}^\infty \mathrm{d}t_1 \dots\! \int\limits_{-\infty}^\infty \mathrm{d}t_n \frac{1}{\braket{{\cal T} S(\infty)}_{0}} \sum_{\substack{3 \text{ perm.} \\ a,b}} \sum_m^n \frac{n!}{(n-m)!} \braket{{\cal T}\hat{O}_a\hat{O}_1}_{0}  \braket{{\cal T}\hat{X}_1\hat{X}_2}_{0} \braket{{\cal T}\hat{O}_2\hat{O}_3}_{0} \dots  \braket{{\cal T}\hat{O}_m\hat{O}_b}_{0} \nonumber\\
&\cdot \sum_k^{n-m} \frac{(n-m)!}{(n-m-k)!} \braket{{\cal T}\hat{O}_c\hat{O}_{m+1}}_{0}  \braket{{\cal T}\hat{X}_{m+1}\hat{X}_{m+2}}_{0} \braket{{\cal T}\hat{O}_{m+2}\hat{O}_{m+3}}_{0} \dots \braket{{\cal T}\hat{O}_{m+k}\hat{O}_d}_{0} \nonumber\\
&\cdot \braket{{\cal T}\hat{O}_{m+k+1} \dots \hat{O}_n}_{0} \braket{{\cal T}\hat{X}_{m+k+1} \dots \hat{X}_n}_{0} \\
&= \sum_{\substack{3 \text{ perm.} \\ a,b}}\frac{1}{\braket{{\cal T} S(\infty)}_{0}} \sum_n^\infty \sum_m^n C^{a,b}_m \sum_k^{n-m} C^{c,d}_k \ V_{n-m-k} = \sum_{\substack{3 \text{ perm.} \\ a,b}} \frac{1}{\braket{{\cal T} S(\infty)}_{0}} \sum_n^\infty V_n \sum_m^\infty C^{a,b}_m \sum_k^\infty C^{c,d}_k \label{eq_resummation_4timecorr_W} \\
&= \sum_{\substack{3 \text{ perm.} \\ a,b}} \braket{{\cal T} \hat{O}_a \hat{O}_b} \braket{{\cal T} \hat{O}_c \hat{O}_d} \,.
\end{align}
The resummation in Eq.~(\ref{eq_resummation_4timecorr_W}) represents the Cauchy product formula for three series followed by an index shift.
Hence, we expressed the full four-time correlator in terms of full two-time correlators.

Now we include the corrections to Wick's theorem and only consider the lowest-order correction $G_4$.
We introduce the correction to the normalization $\braket{{\cal T} S(\infty)}_{0,\text{corr}}$,
\begin{equation}
\frac{1}{\braket{{\cal T} S(\infty)}_{0,F}}=\frac{1}{\braket{{\cal T} S(\infty)}_{0}+\braket{{\cal T} S(\infty)}_{0,\text{corr}}}\approx \frac{1}{\braket{{\cal T} S(\infty)}_{0}} \left( 1 - \frac{\braket{{\cal T} S(\infty)}_{0,\text{corr}}}{\braket{{\cal T} S(\infty)}_{0}} \right)\; .
\end{equation}
With this and Eq.~(\ref{eq_correctiontowick}) we can identify the corrections to the full four-time correlator.
For that we use the following abbreviation to describe on which set of operators we apply Wick's theorem,
\begin{equation}
\text{Wick}(A,\pi_n\backslash B,C) = \braket{{\cal T} \, \prod\limits_{\substack{k \in  A\cup\pi_n\backslash B}} \hat{O}_k}\!{}_{0} \braket{{\cal T} \, \prod\limits_{l \in \pi_n\cup C} \hat{X}_l}\!{}_{0}\; ,
\end{equation}
where $\pi_n=\{1,\dots,n\}$ describes the initial set of operators $\hat{O}_i$ and $\hat{X}_i$. With this notation we keep in mind which additional operators $\hat{O}_i$ we have and which operators $\hat{O}_i$ are missing.
The full four-time correlator with corrections reads
\begin{align}
\braket{{\cal T} \hat{O}_{\text{I}}\hat{O}_{\text{II}}\hat{O}_{\text{III}}\hat{O}_{\text{IV}}}_F =& \sum_{\substack{3 \text{ perm.} \\ a,b,c,d}} \!\! \braket{{\cal T} \hat{O}_a \hat{O}_b} \braket{{\cal T} \hat{O}_c \hat{O}_d}  -\sum_{\substack{3 \text{ perm.} \\ a,b,c,d}} \!\! \braket{{\cal T} \hat{O}_a \hat{O}_b} \braket{{\cal T} \hat{O}_c \hat{O}_d} \frac{\braket{{\cal T} S(\infty)}_{0,\text{corr}}}{\braket{{\cal T} S(\infty)}_{0}}+ G_4(\text{I},\text{II},\text{III},\text{IV}) \nonumber\\
& +\sum_n \frac{(-i)^n}{n!} \!\! \int\limits_{-\infty}^\infty \mathrm{d}t_1 \dots\! \int\limits_{-\infty}^\infty \mathrm{d}t_n \frac{1}{\braket{{\cal T} S(\infty)}_{0}} \left( \sum_{\substack{4 \ \text{perm.} \\ a-d}} \sum_{\substack{\text{perm.} \\ \delta}} G_4(a,b,c,\delta) \text{Wick}(\{d\},\pi_n\backslash\{\delta\}) \right. \nonumber\\
&+ \sum_{\substack{6 \ \text{perm.} \\ a-d}} \sum_{\substack{\text{perm.} \\ \gamma,\delta}} G_4(a,b,\gamma,\delta) \text{Wick}(\{c,d\},\pi_n\backslash\{\gamma,\delta\}) +\sum_{\substack{\text{perm.} \\ \alpha-\delta}} G_4(\alpha,\beta,\gamma,\delta) \text{Wick}(\{a-d\},\pi_n\backslash\{\alpha-\delta\})\nonumber\\
&+\left.  \sum_{\substack{4 \ \text{perm.} \\ a-d}} \sum_{\substack{\text{perm.} \\ \beta,\gamma,\delta}} G_4(a,\beta,\gamma,\delta) \text{Wick}(\{b,c,d\},\pi_n\backslash\{\beta,\gamma,\delta\}) \right) \,.
\end{align}
The summations go over all distinguishable permutations. In addition, the correction to the vacuum diagrams reads
\begin{equation}
\frac{\braket{{\cal T} S(\infty)}_{0,\text{corr}}}{\braket{{\cal T} S(\infty)}_{0}} =
\sum_n \frac{(-i)^n}{n!} \!\! \int\limits_{-\infty}^\infty \mathrm{d}t_1 \dots\! \int\limits_{-\infty}^\infty \mathrm{d}t_n \frac{1}{\braket{{\cal T} S(\infty)}_{0}} 
\sum_{\substack{\text{perm.} \\ \alpha-\delta}} G_4(\alpha,\beta,\gamma,\delta) \text{Wick}(\pi_n\backslash\{\alpha-\delta\}) \,.
\end{equation}
We can do the same for a two-time correlator
\begin{align}
\braket{{\cal T} \hat{O}_a \hat{O}_b}_F =& \braket{{\cal T} \hat{O}_a \hat{O}_b} - \braket{{\cal T} \hat{O}_a \hat{O}_b} \frac{\braket{{\cal T} S(\infty)}_{0,\text{corr}}}{\braket{{\cal T} S(\infty)}_{0}} \nonumber\\
&+\sum_n \frac{(-i)^n}{n!} \!\! \int\limits_{-\infty}^\infty \mathrm{d}t_1 \dots\! \int\limits_{-\infty}^\infty \mathrm{d}t_n \frac{1}{\braket{{\cal T} S(\infty)}_{0}} \left(  \sum_{\substack{\text{perm.} \\ \gamma,\delta}} G_4(a,b,\gamma,\delta) \text{Wick}(\pi_n\backslash\{\gamma,\delta\})\right. \nonumber\\
&+ \left. \sum_{\substack{2 \ \text{perm.} \\ k,l}} \sum_{\substack{\text{perm.} \\ \beta,\gamma,\delta}} G_4(k,\beta,\gamma,\delta) \text{Wick}(l,\pi_n\backslash\{\beta,\gamma,\delta\})+\sum_{\substack{\text{perm.} \\ \alpha-\delta}} G_4(\alpha,\beta,\gamma,\delta) \text{Wick}(a,b,\pi_n\backslash\{\alpha-\delta\} \right) \;.
\end{align}
With these relations we calculate
\begin{equation}
\braket{{\cal T} \hat{O}_{\text{I}}\hat{O}_{\text{II}}\hat{O}_{\text{III}}\hat{O}_{\text{IV}}}_F -\sum_{\substack{3 \text{ perm.} \\ a,b}} \braket{{\cal T}\hat{O}_a\hat{O}_b}_F \braket{{\cal T}\hat{O}_c\hat{O}_d}_F \; .
\end{equation}
We have to compare terms with the same type of $G_4$, because only these terms can cancel each other.
As an example we explain the procedure for $G_4(a,b,\gamma,\delta)$.
Focusing on  $G_4(a,b,\gamma,\delta)$ we obtain
\begin{align}
\sum_n \frac{(-i)^n}{n!} \!\! \int\limits_{-\infty}^\infty \mathrm{d}t_1 \dots\! \int\limits_{-\infty}^\infty \mathrm{d}t_n \frac{1}{\braket{{\cal T} S(\infty)}_{0}}&\left(
\sum_{\substack{6 \ \text{perm.} \\ a-d}} \sum_{\substack{\text{perm.} \\ \gamma,\delta}} G_4(a,b,\gamma,\delta) \text{Wick}(\{c,d\},\pi_n\backslash\{\gamma,\delta\})\right. \nonumber\\
&-\left.\sum_{\substack{6 \ \text{perm.} \\ a-d}} \sum_{\substack{\text{perm.} \\ \gamma,\delta}} G_4(a,b,\gamma,\delta) \text{Wick}(\pi_n\backslash\{\gamma,\delta\}) \braket{{\cal T}\hat{O}_c\hat{O}_d}\right) \;.
\end{align}
In the last term we have to take into account six permutations, since the $G_4(a,b,\gamma,\delta)$ occurs in both two-time correlators.
The summation over the permutations for $\gamma,\delta$ yields a factor $\frac{n!}{(n-2)!}$. Since the first contribution arises for $n=2$ we define $\tilde{n}=n-2$, which yields
\begin{align}
\sum_{\substack{6 \ \text{perm.} \\ a-d}}\!\!(-i)^2 \!\!\int\limits_{-\infty}^\infty \mathrm{d}t_{x_1}\!\!\int\limits_{-\infty}^\infty \mathrm{d}t_{x_2} G_4(a,b,x_1,x_2) \frac{1}{\braket{{\cal T} S(\infty)}_{0}} \sum_{\tilde{n}} \frac{(-i)^{\tilde{n}}}{\tilde{n}!} \!\! \int\limits_{-\infty}^\infty \mathrm{d}t_1 \dots\! \int\limits_{-\infty}^\infty \mathrm{d}t_{\tilde{n}} &\left(\vphantom{\hat{O}_c} \text{Wick}(\{c,d\},\pi_{\tilde{n}},\{x_1,x_2\})\right. \nonumber\\
&-\left. \text{Wick}(\pi_{\tilde{n}},\{x_1,x_2\}) \braket{{\cal T}\hat{O}_c\hat{O}_d} \right) \;. \label{eq_G(a,b,x_1,x_2)}
\end{align}
The possible types of diagrams in these constellations are
\begin{equation}
(\text{I}): \; 
\begin{tikzpicture}[anchor=base,baseline=5pt]
    \coordinate (A) at (0,0);
    \coordinate (B) at (0,0.5);
    \coordinate (C) at (0.5,0);
    \coordinate (D) at (0.5,0.5);
    \draw[line width=1.0pt] (A) -- (D);
    \draw[line width=1.0pt] (B) -- (C);
    \coordinate (L1a) at (1.1,0);
    \coordinate (L2a) at (1.1,0.5);
    \draw[line width=1.0pt, snake it] (C) -- (L1a);
    \draw[line width=1.0pt, snake it] (D) -- (L2a);
    \fill (C) circle (2pt);
    \fill[white] (C) circle (1pt);
    \fill (D) circle (2pt);
    \fill[white] (D) circle (1pt);
    \coordinate (L1b) at (1.4,0);
    \coordinate (L2b) at (1.4,0.5);
    \draw[line width=1.0pt] (L1a) -- (L1b);
    \draw[line width=1.0pt] (L2a) -- (L2b);
    \fill (L1a) circle (2pt);
    \fill[white] (L1a) circle (1pt);
    \fill (L2a) circle (2pt);
    \fill[white] (L2a) circle (1pt);
    \coordinate (L1c) at (1.7,0);
    \coordinate (L2c) at (1.7,0.5);
    \node at (L1c) {\dots};
    \node at (L2c) {\dots};
    \coordinate (L1d) at (1.9,0);
    \coordinate (L2d) at (1.9,0.5);
    \coordinate (L1e) at (2.4,0);
    \coordinate (L2e) at (2.4,0.5);
    \coordinate (L1f) at (3.0,0);
    \coordinate (L2f) at (3.0,0.5);
    \draw[line width=1.0pt,snake it] (L1d) -- (L1e);
    \draw[line width=1.0pt,snake it] (L2d) -- (L2e);
    \draw[line width=1.0pt] (L1e) -- (L1f);
    \draw[line width=1.0pt] (L2e) -- (L2f);
    \fill (L1e) circle (2pt);
    \fill[white] (L1e) circle (1pt);
    \fill (L2e) circle (2pt);
    \fill[white] (L2e) circle (1pt);
\end{tikzpicture} 
\quad (\text{II}): \;
\begin{tikzpicture}[anchor=base,baseline=5pt]
    \coordinate (A) at (0,0);
    \coordinate (B) at (0,0.5);
    \coordinate (C) at (0.5,0);
    \coordinate (D) at (0.5,0.5);
    \draw[line width=1.0pt] (A) -- (D);
    \draw[line width=1.0pt] (B) -- (C);
    \coordinate (R1a) at (1,-0.15);
    \coordinate (R2a) at (1,0.65);
    \draw[line width=1.0pt, snake it] (C) to[out=-45,in=-120] (R1a);
    \draw[line width=1.0pt, snake it] (D) to[out=45,in=120] (R2a);
    \fill (C) circle (2pt);
    \fill[white] (C) circle (1pt);
    \fill (D) circle (2pt);
    \fill[white] (D) circle (1pt);
    \coordinate (R1b) at (1.17,0.05);
    \coordinate (R2b) at (1.17,0.45);
    \draw[line width=1.0pt] (R1a) to[out=30,in=-100] (R1b);
    \draw[line width=1.0pt] (R2a) to[out=-30,in=100] (R2b);
    \fill (R1a) circle (2pt);
    \fill[white] (R1a) circle (1pt);
    \fill (R2a) circle (2pt);
    \fill[white] (R2a) circle (1pt);
    \node at (1.17,0.1) {\vdots};
    \coordinate (E) at (1.7,0.25);
    \coordinate (F) at (2.3,0.25);
    \coordinate (G) at (2.9,0.25);
    \coordinate (H) at (3.2,0.25);
    \coordinate (I) at (3.5,0.25);
    \draw[line width=1.0pt] (E) -- (F);
    \draw[line width=1.0pt, snake it] (F) -- (G);
    \draw[line width=1.0pt] (G) -- (H);
    \fill (F) circle (2pt);
    \fill[white] (F) circle (1pt);
    \fill (G) circle (2pt);
    \fill[white] (G) circle (1pt);
    \node at (I) {\dots};
    \coordinate (J) at (3.8,0.25);
    \coordinate (K) at (4.3,0.25);
    \coordinate (L) at (4.9,0.25);
    \draw[line width=1.0pt, snake it] (J) -- (K);
    \draw[line width=1.0pt] (K) -- (L);
    \fill (K) circle (2pt);
    \fill[white] (K) circle (1pt);
\end{tikzpicture} \;,
\end{equation}
multiplied by an appropriate vacuum diagram. The cross represents $G_4$. Both kinds appear in the first term. However, all contributions in the second term are of the form (II).
We define the following abbreviations that describe the leg- and ring-type structures in the above diagrams:
\begin{align}
L_m^{x_1,a}&= (-i)^m \!\!\!\int\limits_{-\infty}^\infty \mathrm{d}t_1 \dots\! \int\limits_{-\infty}^\infty \mathrm{d}t_m \braket{{\cal T}\hat{X}_{x_1}\hat{X}_{1}}_{0} \braket{{\cal T}\hat{O}_{1}\hat{O}_{2}}_{0} \braket{{\cal T}\hat{X}_{2}\hat{X}_{3}}_{0} \dots \braket{{\cal T}\hat{O}_{m}\hat{O}_{a}}_{0} \;,\; L_0^{x_1,a}=0 \\
R_m^{x_1,x_2} &=(-i)^m \!\!\!\int\limits_{-\infty}^\infty \mathrm{d}t_1 \dots\! \int\limits_{-\infty}^\infty \mathrm{d}t_m \braket{{\cal T}\hat{X}_{x_1}\hat{X}_{1}}_{0} \braket{{\cal T}\hat{O}_{1}\hat{O}_{2}}_{0} \braket{{\cal T}\hat{X}_{2}\hat{X}_{3}}_{0} \dots \braket{{\cal T}\hat{X}_{m}\hat{X}_{x_2}}_{0}
\; .
\end{align}
Now we proceed analogously with Eq.~(\ref{eq_G(a,b,x_1,x_2)}) and identify similar structures, get combinational factors, and do the resummation using the Cauchy product formula. It turns out that the terms of type (II) fully cancel out. So we are left with
\begin{equation}
\sum_{\substack{6 \ \text{perm.} \\ a-d}}\!\!(-i)^2 \!\!\int\limits_{-\infty}^\infty \mathrm{d}t_{x_1}\!\!\int\limits_{-\infty}^\infty \mathrm{d}t_{x_2} G_4(a,b,x_1,x_2) \sum_l^{\infty} \sum_k^{\infty} L_l^{x_1,c} L_k^{x_2,d} \;.
\end{equation}
Repeating this procedure for all kinds of $G_4$ terms, we find
\begin{align}
\braket{{\cal T} \hat{O}_{\text{I}}\hat{O}_{\text{II}}\hat{O}_{\text{III}}\hat{O}_{\text{IV}}}_F =& \sum_{\substack{3 \text{ perm.} \\ a,b}} \braket{{\cal T}\hat{O}_a\hat{O}_b}_F \braket{{\cal T}\hat{O}_c\hat{O}_d}_F +G_4(\text{I},\text{II},\text{III},\text{IV})
-i\!\sum_{\substack{4 \ \text{perm.} \\ a-d}}\!\!\!\int\limits_{-\infty}^\infty \mathrm{d}t_{x_1} G_4(a,b,c,x_1) \sum_k^{\infty} L_k^{x_1,d} \nonumber\\
&- \!\sum_{\substack{6 \ \text{perm.} \\ a-d}}\!\!\!\int\limits_{-\infty}^\infty \mathrm{d}t_{x_1}\!\!\int\limits_{-\infty}^\infty \mathrm{d}t_{x_2} G_4(a,b,x_1,x_2) \sum_l^{\infty} \sum_k^{\infty} L_l^{x_1,c} L_k^{x_2,d} \nonumber\\
&+i\!\sum_{\substack{4 \ \text{perm.} \\ a-d}}\!\!\!\int\limits_{-\infty}^\infty \mathrm{d}t_{x_1}\!\!\int\limits_{-\infty}^\infty \mathrm{d}t_{x_2} \!\!\int\limits_{-\infty}^\infty \mathrm{d}t_{x_3}G_4(a,x_1,x_2,x_3) \sum_l^{\infty} \sum_k^{\infty}\sum_m^{\infty} L_l^{x_1,b} L_k^{x_2,c}L_m^{x_3,d} \nonumber\\
&+\int\limits_{-\infty}^\infty \mathrm{d}t_{x_1}\!\!\int\limits_{-\infty}^\infty \mathrm{d}t_{x_2} \!\!\int\limits_{-\infty}^\infty \mathrm{d}t_{x_3}\!\!\int\limits_{-\infty}^\infty \mathrm{d}t_{x_4}G_4(x_1,x_2,x_3,x_4) \sum_l^{\infty} \sum_k^{\infty}\sum_m^{\infty}\sum_n^{\infty} L_l^{x_1,a} L_k^{x_2,b}L_m^{x_3,c} L_m^{x_4,d} \,.
\end{align}
We define a diagrammatic representation for these corrections,
\begin{equation}
\begin{tikzpicture}[anchor=base,baseline=5pt]
    \coordinate (A) at (0,0);
    \coordinate (B) at (0,0.5);
    \coordinate (C) at (0.5,0);
    \coordinate (D) at (0.5,0.5);
    \draw[line width=2.0pt] (A) -- (D);
    \draw[line width=2.0pt] (B) -- (C);
\end{tikzpicture}
=
\begin{tikzpicture}[anchor=base,baseline=5pt]
    \coordinate (A) at (0,0);
    \coordinate (B) at (0,0.5);
    \coordinate (C) at (0.5,0);
    \coordinate (D) at (0.5,0.5);
    \draw[line width=1.0pt] (A) -- (D);
    \draw[line width=1.0pt] (B) -- (C);
\end{tikzpicture}
+
\begin{tikzpicture}[anchor=base,baseline=5pt]
    \coordinate (A) at (0,0);
    \coordinate (B) at (0,0.5);
    \coordinate (C) at (0.5,0);
    \coordinate (D) at (0.5,0.5);
    \draw[line width=1.0pt] (A) -- (D);
    \draw[line width=1.0pt] (B) -- (C);
    \coordinate (L1a) at (1.1,0);
    \draw[line width=1.0pt, snake it] (C) -- (L1a);
    \fill (C) circle (2pt);
    \fill[white] (C) circle (1pt);
    \coordinate (L1b) at (1.7,0);
    \draw[line width=1.0pt] (L1a) -- (L1b);
    \fill (L1a) circle (2pt);
    \fill[white] (L1a) circle (1pt);
\end{tikzpicture}
+
\begin{tikzpicture}[anchor=base,baseline=5pt]
    \coordinate (A) at (0,0);
    \coordinate (B) at (0,0.5);
    \coordinate (C) at (0.5,0);
    \coordinate (D) at (0.5,0.5);
    \draw[line width=1.0pt] (A) -- (D);
    \draw[line width=1.0pt] (B) -- (C);
    \coordinate (L1a) at (1.1,0);
    \coordinate (L2a) at (1.1,0.5);
    \draw[line width=1.0pt, snake it] (C) -- (L1a);
    \draw[line width=1.0pt, snake it] (D) -- (L2a);
    \fill (C) circle (2pt);
    \fill[white] (C) circle (1pt);
    \fill (D) circle (2pt);
    \fill[white] (D) circle (1pt);
    \coordinate (L1b) at (1.7,0);
    \coordinate (L2b) at (1.7,0.5);
    \draw[line width=1.0pt] (L1a) -- (L1b);
    \draw[line width=1.0pt] (L2a) -- (L2b);
    \fill (L1a) circle (2pt);
    \fill[white] (L1a) circle (1pt);
    \fill (L2a) circle (2pt);
    \fill[white] (L2a) circle (1pt);
\end{tikzpicture}
+
\begin{tikzpicture}[anchor=base,baseline=5pt]
    \coordinate (A) at (0,0);
    \coordinate (B) at (0,0.5);
    \coordinate (C) at (0.5,0);
    \coordinate (D) at (0.5,0.5);
    \draw[line width=1.0pt] (A) -- (D);
    \draw[line width=1.0pt] (B) -- (C);
    \coordinate (L1a) at (1.1,0);
    \draw[line width=1.0pt, snake it] (C) -- (L1a);
    \fill (C) circle (2pt);
    \fill[white] (C) circle (1pt);
    \coordinate (L1b) at (1.7,0);
    \draw[line width=1.0pt] (L1a) -- (L1b);
    \fill (L1a) circle (2pt);
    \fill[white] (L1a) circle (1pt);
    \coordinate (L1c) at (2.3,0);
    \draw[line width=1.0pt, snake it] (L1b) -- (L1c);
    \fill (L1b) circle (2pt);
    \fill[white] (L1b) circle (1pt);
    \coordinate (L1d) at (2.9,0);
    \draw[line width=1.0pt] (L1c) -- (L1d);
    \fill (L1c) circle (2pt);
    \fill[white] (L1c) circle (1pt);
\end{tikzpicture}
+\dots \;,
\end{equation}
and are left with
\begin{equation}
\braket{{\cal T} \hat{O}_{\text{I}}\hat{O}_{\text{II}}\hat{O}_{\text{III}}\hat{O}_{\text{IV}}}_F = \sum_{\substack{3 \text{ perm.} \\ a,b,c,d}} \braket{{\cal T}\hat{O}_a\hat{O}_b}_F \braket{{\cal T}\hat{O}_c\hat{O}_d}_F 
+ \;
\begin{tikzpicture}[anchor=base,baseline=5pt]
    \coordinate (A) at (0,0);
    \coordinate (B) at (0,0.5);
    \coordinate (C) at (0.5,0);
    \coordinate (D) at (0.5,0.5);
    \draw[line width=2.0pt] (A) -- (D);
    \draw[line width=2.0pt] (B) -- (C);
\end{tikzpicture} \,.
\end{equation}
With the above it is easy to derive the equation
\begin{align}
\braket{{\cal T} \hat{O}_{\text{I}} \hat{O}_{\text{II}}}_F =& \braket{{\cal T} \hat{O}_{\text{I}} \hat{O}_{\text{II}}}
- \int\limits_{-\infty}^\infty \mathrm{d}t_{x_1}\!\!\int\limits_{-\infty}^\infty \mathrm{d}t_{x_2} G_4(\text{I},\text{II},x_1,x_2)  \sum_k^{\infty} R_k^{x_1,x_2} \nonumber\\
&+i\!\sum_{\substack{2 \ \text{perm.} \\ a,b}}\!\!\!\int\limits_{-\infty}^\infty \mathrm{d}t_{x_1}\!\!\int\limits_{-\infty}^\infty \mathrm{d}t_{x_2} \!\!\int\limits_{-\infty}^\infty \mathrm{d}t_{x_3}G_4(a,x_1,x_2,x_3) \sum_l^{\infty} \sum_k^{\infty} R_k^{x_1,x_2} L_l^{x_3,b} \nonumber\\
&+\int\limits_{-\infty}^\infty \mathrm{d}t_{x_1}\!\!\int\limits_{-\infty}^\infty \mathrm{d}t_{x_2} \!\!\int\limits_{-\infty}^\infty \mathrm{d}t_{x_3}\!\!\int\limits_{-\infty}^\infty \mathrm{d}t_{x_4}G_4(x_1,x_2,x_3,x_4) \sum_l^{\infty} \sum_k^{\infty}\sum_m^{\infty}R_k^{x_1,x_2} L_l^{x_3,\text{I}} L_m^{x_4,\text{II}} \\
=& \braket{{\cal T} \hat{O}_{\text{I}} \hat{O}_{\text{II}}}
+
\begin{tikzpicture}[anchor=base,baseline=5pt]
    \coordinate (A) at (0,0);
    \coordinate (B) at (0,0.5);
    \coordinate (C) at (0.5,0);
    \coordinate (D) at (0.5,0.5);
    \draw[line width=2.0pt] (A) -- (D);
    \draw[line width=2.0pt] (B) -- (C);
    \draw[line width=1.0pt,snake it] (D) to[out=-45,in=45] (C);
    \fill (C) circle (2pt);
    \fill[white] (C) circle (1pt);
    \fill (D) circle (2pt);
    \fill[white] (D) circle (1pt);
\end{tikzpicture} \;.
\end{align}

\end{widetext}


\end{document}